\documentclass[acmlarge, screen]{acmart}

\usepackage{booktabs} 
\usepackage{graphicx} 
\usepackage{verbatim}

\def\plaintitle{FarSense: Pushing the Range Limit of WiFi-based Respiration Sensing with CSI Ratio of Two Antennas}

\setcopyright{acmcopyright}

\acmJournal{IMWUT}
\acmYear{2019}
\acmVolume{3}
\acmNumber{3}
\acmArticle{121}
\acmMonth{9}
\acmPrice{15.00}

\acmDOI{10.1145/3351279}

\begin{document}
\title{\plaintitle}
\author{Youwei Zeng}
\affiliation{ 
	\department{Key Laboratory of High Confidence Software Technologies (Ministry of Education), School of Electronics Engineering and Computer Science}
	\institution{Peking University}
	\city{Beijing}
	\country{China}
}
\email{ywzeng@pku.edu.cn}

\author{Dan Wu}
\affiliation{ 
	\department{Key Laboratory of High Confidence Software Technologies (Ministry of Education), School of Electronics Engineering and Computer Science}
	\institution{Peking University}
	\city{Beijing}
	\country{China}
}
\email{dan@pku.edu.cn}

\author{Jie Xiong}
\affiliation{ 
	\department{College of Information and Computer Sciences}
	\institution{University of Massachusetts}
	\city{Amherst}
	\country{USA}
}
\email{jxiong@cs.umass.edu}

\author{Enze Yi}
\affiliation{ 
	\department{Key Laboratory of High Confidence Software Technologies (Ministry of Education), School of Electronics Engineering and Computer Science}
	\institution{Peking University}
	\city{Beijing}
	\country{China}
}
\email{yienze_cs@pku.edu.cn}

\author{Ruiyang Gao}
\affiliation{ 
	\department{Key Laboratory of High Confidence Software Technologies (Ministry of Education), School of Electronics Engineering and Computer Science}
	\institution{Peking University}
	\city{Beijing}
	\country{China}
}
\email{gry@pku.edu.cn}

\author{Daqing Zhang}
\authornote{This is the corresponding author}
\affiliation{ 
	\department{Key Laboratory of High Confidence Software Technologies (Ministry of Education), School of Electronics Engineering and Computer Science}
	\institution{Peking University}
	\city{Beijing}
	\country{China}
}
\affiliation{ 
	\department{Institut Mines}
	\institution{Telecom SudParis}
	\city{Evry}
	\country{France}
}
\email{dqzhang@sei.pku.edu.cn}

\begin{abstract}
	
The past few years have witnessed the great potential of exploiting channel state information retrieved from commodity WiFi devices for respiration monitoring.
However, existing approaches only work when the target is close to the WiFi transceivers and the performance degrades significantly when the target is far away.
On the other hand, most home environments only have one WiFi access point and it may not be located in the same room as the target. This sensing range constraint greatly limits the application of the proposed approaches in real life.

This paper presents FarSense--the first real-time system that can reliably monitor human respiration when the target is far away from the WiFi transceiver pair. FarSense works well even when one of the transceivers is located in another room, moving a big step towards real-life deployment.  
We propose two novel schemes to achieve this goal:
(1) Instead of applying the raw CSI readings of individual antenna for sensing, we employ the ratio of CSI readings from two antennas, whose noise is mostly canceled out by the division operation to significantly increase the sensing range;
(2) The division operation further enables us to utilize the phase information which is not usable with one single antenna for sensing. The orthogonal amplitude and phase are elaborately combined to address the "blind spots" issue and further increase the sensing range. Extensive experiments show that FarSense is able to accurately monitor human respiration even when the target is 8 meters away from the transceiver pair, increasing the sensing range by more than 100\%.\footnotemark[1]
\footnotetext[1]{Different from radar with transmitter and receiver at the same location, the WiFi transmitter and receiver are separated physically. Here, the distance between a target and a transceiver pair is defined as the average distance from the target to the transmitter and the receiver.}
We believe this is the first system to enable through-wall respiration sensing with commodity WiFi devices and the proposed method could also benefit other sensing applications.   

\end{abstract}



\maketitle

\renewcommand{\shortauthors}{Zeng et al.}
\renewcommand{\shorttitle}{FarSense: Pushing the Range Limit of WiFi-based Respiration Sensing with CSI Ratio of...}

\section{Introduction}
\label{sec:introduction}

\begin{figure*}[t]
	\begin{minipage}[t]{0.7\linewidth}
		\centering
		\includegraphics[width=1\textwidth]{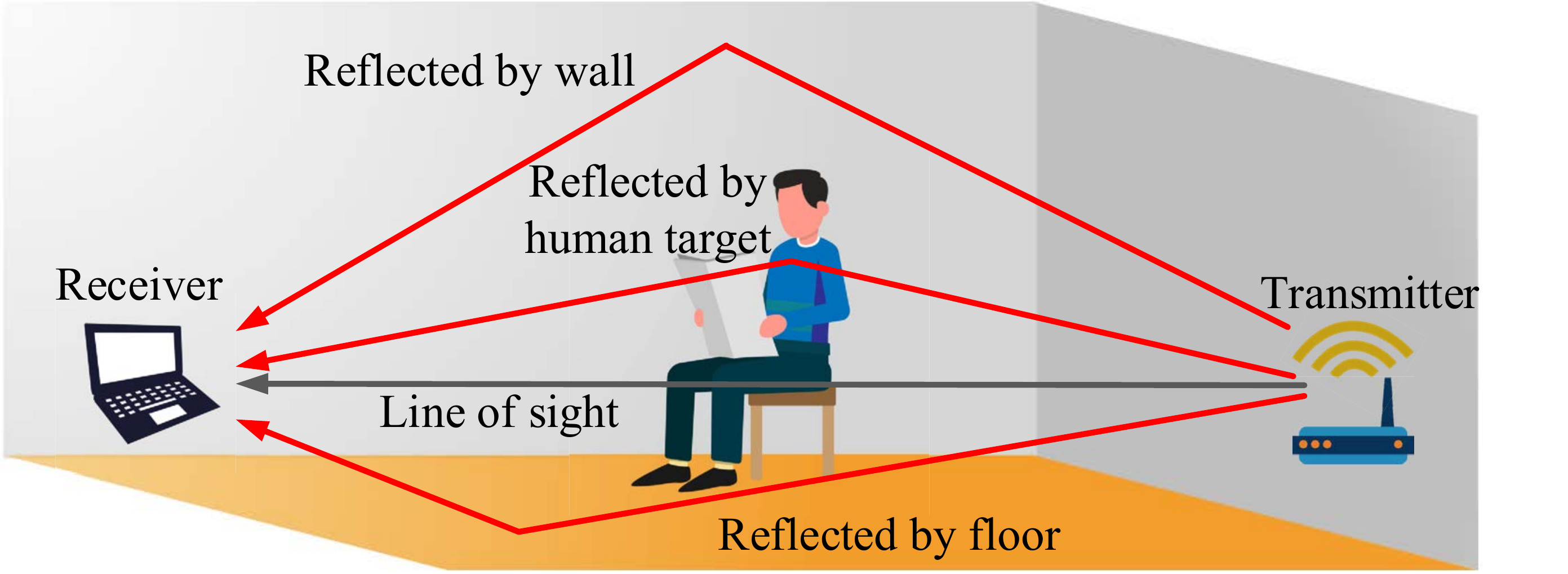}
		\caption{
			In an indoor environment, WiFi signals propagate through multiple paths before arriving at the receiver, thus carrying information about the environment.
		}
		\label{fig:sensing}
	\end{minipage}
\end{figure*}

Wireless technologies have achieved a great success in data communication in the last two decades. Wi-Fi and cellular networks enable us to be seamlessly connected to the Internet, changing our lives in every aspect. In the indoor environment, Wi-Fi is the preferred choice due to its high rate and low cost.
According to \cite{liverman2016hybrid}, there are approximately five billion devices communicating over WiFi networks worldwide in 2016 and this number is expected to increase to thirty billion by 2020. In the last few years, Wi-Fi signals are creatively exploited for contactless human activity sensing.
As shown in Fig.~\ref{fig:sensing}, in an indoor environment, WiFi signals propagate through direct path and also reflection paths, bouncing off the objects in the environment before arriving at the receiver. With careful signal processing on the reflected signal, we can actually obtain a lot of useful information about the objects such as the direction of the signal which can be utilized for localization. When the object is a human target, we can employ the signal variations to infer the target's gestures and activities, enabling human sensing without a device attached to the human body. 

In the last few years, we have seen a lot of emerging WiFi-based human sensing applications, ranging from coarse-grained activity recognition~\cite{shahzad2018augmenting, wang2015understanding, palipana2018falldefi, yu2018qgesture, wang2017rt, ma2018signfi, wang2016gait, ali2015keystroke, li2016wifinger, zhang2019towards}, indoor localization/tracking \cite{li2017indotrack, gong2018sifi, qian2017widar, wu2016widir}, intrusion/motion detection \cite{wu2015non, xin2018freesense, li2018training, li2017ar} to fine-gained respiration monitoring \cite{zhang2018fresnel, zeng2018fullbreathe, wang2016human, wang2017phasebeat, liu2015tracking, liu2014wi, zhang2017toward, liu2016contactless, wang2017tensorbeat, hillyard2018experience, wu2017device, niu2018boosting}.
As one of the most important applications, respiration monitoring attracts lots of attention since respiration is an important health metric used for tracking diseases in many areas, such as sleep, pulmonology and cardiology.
The features extracted from human respiration also provide useful insights about the psychological state of an individual \cite{kunik2005surprisingly, ley1994breathing}.
Among these features, respiration rate is particularly important \cite{elliott2016respiratory} and needs to be carefully observed.
For example, an abnormal respiration rate, either too high (tachypnea), too low (bradypnea), or absent (apnea), is a sensitive indicator of physiological distress that requires immediate clinical intervention \cite{wang2016human}.

Traditional approaches employ wearables or cameras for respiration monitoring. However, the elderly are usually reluctant to wear wearables and the camera-based approaches raise severe privacy concerns.
The latest research in this area explore the possibility of applying pervasive WiFi signals for respiration monitoring without attaching a device on the target.
Multiple efforts have been made for accurate respiration monitoring leveraging either CSI amplitude of a single antenna or phase difference between two antennas available at commodity WiFi hardware~\cite{liu2015tracking, liu2014wi, wu2015non, wang2016human, zhang2017wicare, wang2017phasebeat, wang2017tensorbeat}.   
However, the problem with the existing work is that the sensing range is still very limited, and the sensing devices are required to be placed close to the target. While the communication range of WiFi can be tens of meters, the sensing range is limited to 2-4 meters. The main reason is that wireless sensing relies on weak reflected signal and the subtle respiration-induced signal variation can be easily buried in noise. Furthermore, the existing works face the “blind-spot” issue that respiration cannot be effectively detected at certain locations even when the target is close to the sensing devices. The short sensing range and “blind-spot” constraints greatly limit the real-life application of the existing approaches.

In this paper, for the first time, we present novel solutions to push the respiration sensing range from the current 2-4 meters to house level~(8-9 meters) with commodity WiFi devices, bridging the gap between lab prototype and real-life deployment. The key idea is that we employ the widely available two antennas at commodity WiFi AP for performance boosting. With just two antennas, we construct a new metric -- the ratio of CSI readings of two antennas.
With this division operation between two antennas, most of the noise in the original CSI amplitude and the time-varying phase offset are canceled out.
The CSI ratio of two antennas obtained is much more noise-free and sensitive compared to the original CSI reading from a single antenna when sensing subtle movements. Another big advantage of this "CSI ratio" is that phase information can now be utilized together with the amplitude for sensing. Note that previously people usually only employ the CSI amplitude for sensing because the CSI phase is not stable due to the lack of tight time synchronization between transmitter and receiver. The phase of the ratio is stable as the time-varying random offsets are the same at both antennas and are thus canceled by the division operation.
We further combine the phase and amplitude of the CSI ratio which are complementary to each other in terms of sensing capability to remove the "blind spots" reported in \cite{wang2016human}.
With the two proposed techniques, we significantly extend the sensing range to 8 meters with commodity WiFi device while still keep the 100\% detection rate, outperforming the state-of-the-art WiFi-based approaches. 
What’s more, in this work, we propose the CSI-ratio model that establishes the relationship between the target movement and CSI ratio changes which lays the foundation to guide wireless sensing with CSI ratio.
We believe this general CSI-ratio model will benefit not just respiration sensing but a lot of other CSI-based sensing applications as well. We also believe this general signal ratio method can be applied to other wireless technologies such as RFID (Radio Frequency IDentification), LTE (Long Term Evolution) \cite{pecoraro2017lte} and LoRa (Long Range) \cite{liando2019known} to increase the sensing range and performance.
To validate the effectiveness and robustness of the proposed system FarSense, we conduct comprehensive experiments with different subjects in different environments by varying the settings below:
\begin{itemize}
\item the subject is located at different distances to the transceivers;
\item the challenging scenario when the WiFi transmitter is placed in a different room from the receiver with a wall in between, and the subject is located either in the transmitter side or in the receiver side;
\item another challenging scenario when both transceivers are mounted on the ceiling, far away from the subject.
\end{itemize}
The demo video of FarSense that works in all three scenarios above is also submitted as an accompanying material to show the effectiveness and robustness of the proposed techniques in real-world settings.

The main contributions of the work can be summarized as follows:
\begin{enumerate}
	\item We propose to employ CSI ratio rather than raw CSI for sensing. We develop the CSI-ratio model that establishes the relationship between human's movement and CSI ratio changes which lays the foundation to guide fine-grained sensing. We believe the general CSI-ratio model will benefit not just respiration sensing but a lot of other sensing applications.
	\item We apply the CSI-ratio model for respiration sensing and elaborately combine the amplitude and phase of CSI ratio to address the "blind spots" issue and further increase the sensing range.
	\item We design and implement FarSense on commodity WiFi devices. The sensing range is increased from the current state-of-the-art 2-4 meters to 8-9 meters. For the first time, FarSense is able to enable through-wall respiration sensing with commodity WiFi hardware, moving one step further towards real-life deployment.
\end{enumerate}

The rest of this paper is organized as follows.
Sec.~\ref{sec:related_work} surveys the related work.
Sec.~\ref{sec:empirical_study} employs benchmark experiments to demonstrate the ratio of CSI between two antennas has higher sensitivity than CSI of one antenna for fine-grained sensing.
Sec.~\ref{sec:model} presents the CSI-ratio model and its verification.
Sec.~\ref{sec:pattern_extraction} shows how to effectively extract respiration pattern from the CSI ratio.
Sec.~\ref{sec:system} presents the detailed design and implementation of the prototype system FarSense.
Sec.~\ref{sec:evaluation} presents the experimental setup and evaluation results.
Sec.~\ref{sec:discussions} discusses the limitations and opportunities followed by a conclusion in Sec.~\ref{sec:conclusion}.
\section{Related Work}
\label{sec:related_work}

In this section, we discuss the most related work in respiration sensing with RF signals which can be broadly grouped into two following categories.	

\textbf{Radar-based respiration sensing.}
These approaches can be mainly divided into three categories according to the technologies they are using.
\begin{enumerate}
	\item Continuous-wave (CW) Doppler radar.
	CW Doppler radar technique is based on detecting the reflected frequency echoes to the chest wall motion during respiration \cite{droitcour2004range, droitcour2006non}.
	It has the advantages of low power consumption and simple radio architecture.
	However, it suffers from the clutter noise and multipath reflection so that vital sign signals may not be differentiable from other received frequency echoes.
	
	\item Ultra-wideband (UWB) pulse radar.
	The basic operation of UWB pulse radar is to send a train of pulses towards the target and then the received signal can be visualized in the frequency domain \cite{immoreev2008uwb, lazaro2010analysis}.
	To get an accurate estimation of vital signs, the delay profile in time domain is extracted using IFFT.
	With a bandwidth of 1-2\,GHz, the UWB pulse radar can eliminate interference caused by reflection from other objects and multipath reflection \cite{li2013review}.
	However, such a wide bandwidth requires precise control on the pulse width and the radar peak signal intensity \cite{lai2011wireless}, thus increasing the hardware requirements and system complexity.
		 
	\item Frequency-modulated continuous-wave (FMCW) radar.
	FMCW radar radiates continuous transmission power like a CW radar but linearly increases operating frequency of the transmitted signal during the measurement within a wide bandwidth (for example, 1.79\,GHz in \cite{adib2015smart}). By comparing the frequency of the received signal bounced off human body to the transmission signal, FMCW radar can directly measure the distance of the reflection body from the device. Vital-Radio \cite{adib2015smart} separates signals from different users based on their reflection time and analyze the signals from each user to measure his/her respiration rate and heart rate using FFT.
	EQ-Radio \cite{zhao2016emotion} proposes a new algorithm to extract individual heartbeats from the wireless signal and recognizes a person's emotional state based on his/her respiration signal and heartbeat segmentation.
	All these works are based on FMCW signal, not WiFi (CSI). FMCW hardware employs a much wider bandwidth (1.79\,GHz in \cite{adib2015smart}) which is not available with cheap commodity WiFi hardware (the bandwidth of WiFi is 20/40\,MHz). Further, the costs of FMCW hardware are usually much higher which make these solutions less practical for everyday home usage.
	In this work, we would like to employ the commodity WiFi hardware already pervasively deployed in our home environment for respiration sensing.
\end{enumerate}

\textbf{WiFi-based respiration sensing.}
These WiFi-based approaches can be divided into two categories based on which part of the information from the complex-valued CSI they are using.
\begin{enumerate}
	\item Respiration sensing with amplitude information.
	Most existing approaches employ CSI amplitude information for respiration sensing.
	CSI amplitude is also widely used in other WiFi-based human sensing applications, even though it contains relatively large noise caused by power amplifier uncertainty~(PAU) and environmental noise \cite{zhuo2017perceiving, wang2015understanding, yu2018qgesture}.
	CSI amplitude has explicit mathematical and physical relationship with the target's movements, as demonstrated in~\cite{wang2015understanding}.
	Wi-Sleep~\cite{liu2014wi} is the first sleep monitoring system that extracts rhythmical patterns caused by respiration from WiFi signals. The performance is further improved in~\cite{liu2016contactless} by considering the sleep postures and abnormal respiration patterns.
	Liu~\emph{et al.} \cite{liu2015tracking} have shown to track human heart rate by analyzing power spectral density (PSD) of CSI amplitude during sleep. Wu \emph{et al.} \cite{wu2015non} have demonstrated that respiration information can be obtained no matter the target is lying or standing.
	These works leverage the amplitude information of CSI and the "blind spots" issue still exists.
	A recent work~\cite{wang2016human} introduces the Fresnel zone theory to explain why the blind spots occur.
	
	\item Respiration sensing with phase information.
	The phase information of the CSI readings retrieved from commodity WiFi devices is especially noisy due to the presence of sampling frequency offset~(SFO), carrier frequency offset~(CFO) and packet detection delay~(PDD) \cite{kotaru2015spotfi, vasisht2016decimeter, xie2015precise, yu2018qgesture}, which make it not directly usable for sensing.
	Multiple approaches \cite{kotaru2015spotfi, zhu2017calibrating, yu2018qgesture} have been proposed to calibrate out the CSI phase offset, including SFO and CFO calibration.
	However, as shown in \cite{zeng2018fullbreathe}, these proposed approaches still fail to monitor fine-grained millimeter-level chest motion caused by respiration.
	
	Thus, researchers employ the phase difference between two antennas instead to cancel out the time-varying phase offset since they are the same across two antennas on a same WiFi card \cite{kotaru2015spotfi, qian2017inferring, li2017indotrack}.
	In~\cite{wang2017tensorbeat, wang2017phasebeat}, researchers demonstrated that phase difference can be utilized for respiration monitoring. However, the phase difference between two antennas can be constructive or destructive depending on their in-phase or out-phase relationship. When they are in phase, "blind spots" occur and the sensing performance is poor. The latest work~\cite{zeng2018fullbreathe} demonstrated the complementary property between phase and amplitude for sensing and employ both amplitude and phase to remove the amount of "blind spots". However, it still fails to address the fundamental "small range" issue for WiFi-based respiration sensing.
\end{enumerate}

Differently, we propose to employ the ratio of CSI readings from two antennas as a new signal for the first time for respiration sensing. With this simple division operation between two antennas, most of the noise in the original CSI amplitude and the time-varying phase offset are canceled out. With the random phase offset removed, we are now able to retrieve stable phase information for sensing. We further combine the phase and amplitude of the CSI ratio which are complementary to each other in terms of sensing capability to remove the "blind spots". These two novel schemes enable FarSense to achieve robust sensing performance and a much larger sensing range, moving one big step towards practical application of WiFi-based respiration sensing in real life.
\section{Empirical Study}
\label{sec:empirical_study}

In this section, we conduct the proof-of-concept experiments using commodity WiFi devices to see if the ratio of CSI between two antennas has higher sensitivity than the CSI of one antenna.
Note that the amplitude of the CSI ratio equals to the ratio of two CSI amplitudes at the two antennas while the phase of the CSI ratio equals to the CSI phase difference between the two antennas.
Since the CSI phase of one antenna is not directly usable for sensing, here, we focus on comparing the amplitude information, i.e., the CSI amplitude of one antenna which was widely used for sensing with the ratio of CSI amplitude between two antennas proposed by us.

\begin{figure*}[t]
	\begin{minipage}[t]{0.4\linewidth}
		\centering
		\includegraphics[width=1\textwidth]{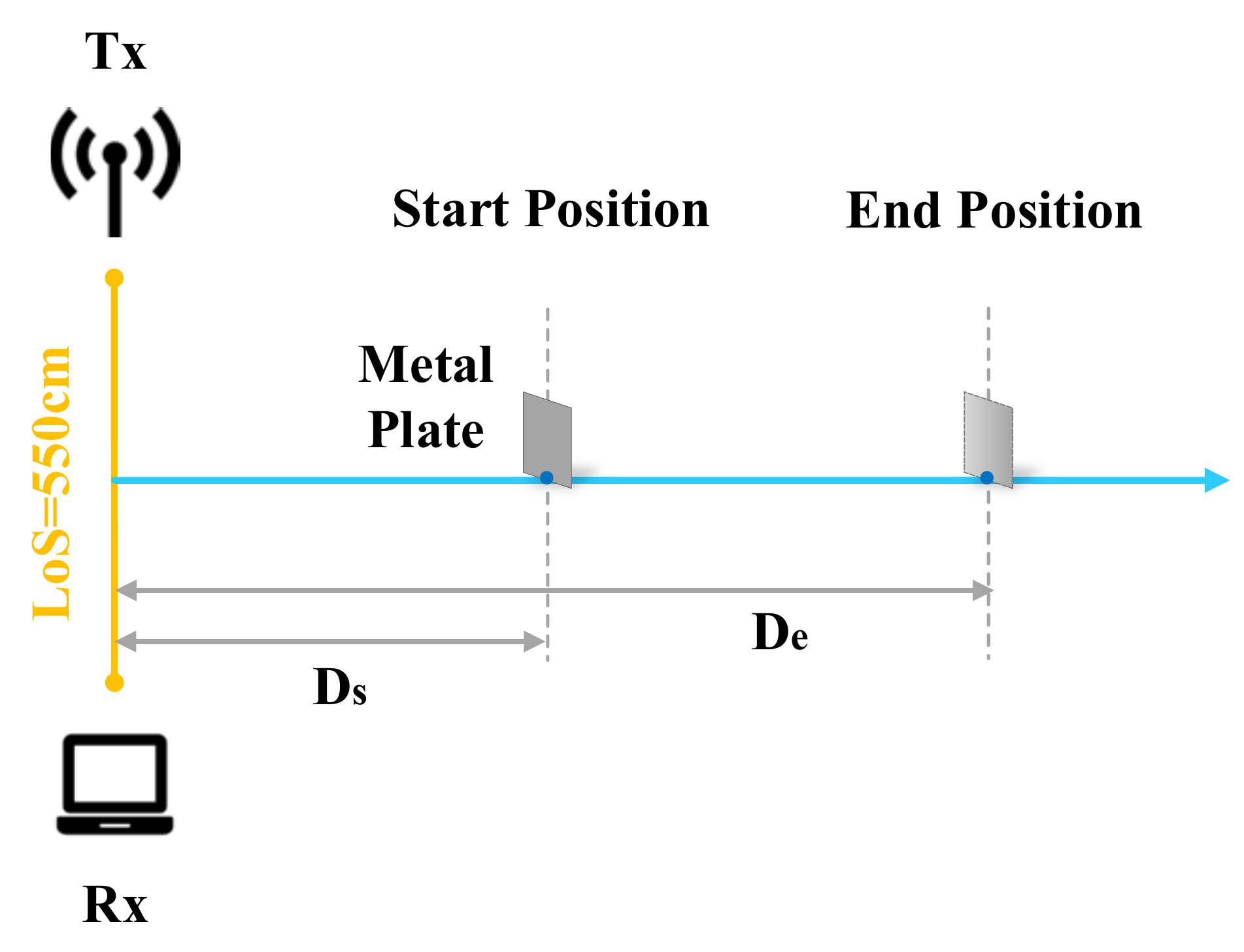}
		\caption{
			Experimental settings: the metal plate moves from the start position to the end position.
		}
		\label{fig:es_setting}
	\end{minipage}
\end{figure*}

\textbf{Experimental Settings.}
As shown in Fig.~\ref{fig:es_setting}, we move a square metal plate with a side length of 15\ cm along the perpendicular bisector of the transceivers with a LoS length of 5.5\ m in an empty room. The transceivers and the metal plate are placed at the same height (1.2\,m).   
Since the flat metal plate can serve as a perfect reflector for radio waves, there is only one dominating reflection path from the plate.
For the experiment, we move the metal plate along the perpendicular bisector of the transceiver pair in the range of 5\,m to 7\,m. We record the ground truth of the start position $D_s$ and end position $D_e$.
Following simple geometry, the ground truth of the reflection path length change is $2\sqrt{(\frac{LoS}{2})^{2}+D_e^{2}}-2\sqrt{(\frac{LoS}{2})^{2}+D_s^{2}}$.
We also vary the LoS path length to see the effect.

\begin{figure*}[t]
	\begin{minipage}[t]{0.99\linewidth}
		\centering
		\includegraphics[width=1\textwidth]{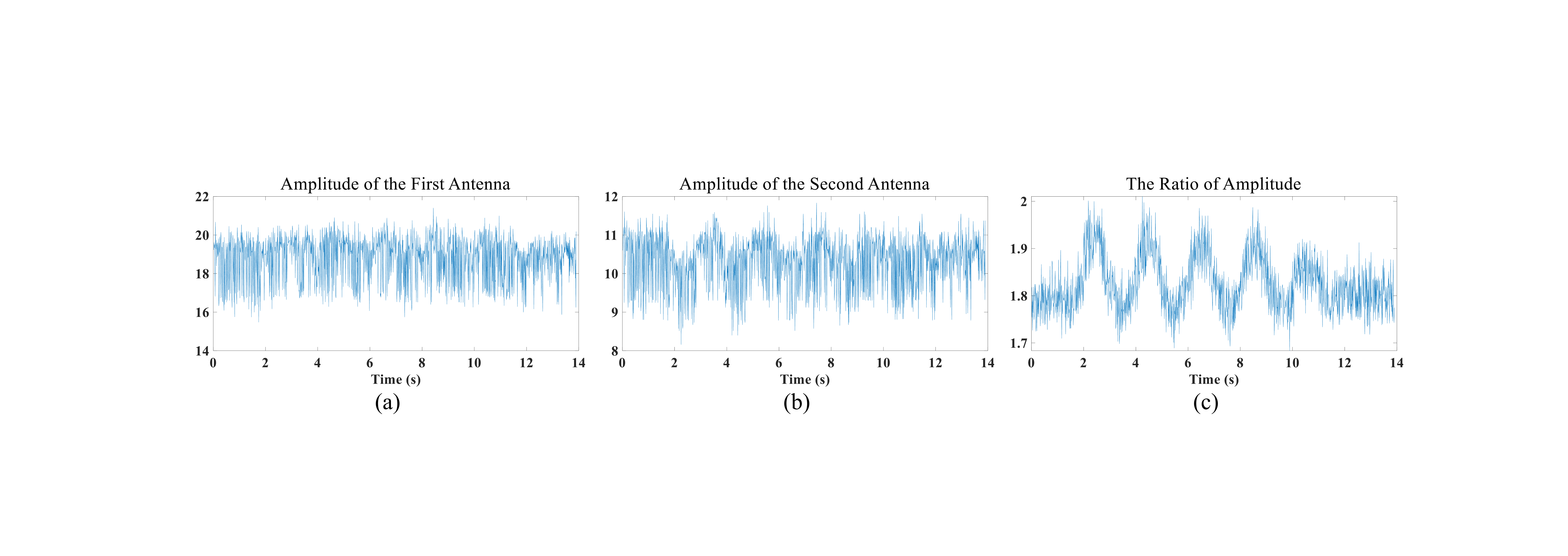}
		\caption{
			Comparison of three amplitude waveforms when a metal plate moves further away.
			Obviously, the ratio of amplitude outperforms the other two raw amplitude waveforms for its clear fluctuation caused by the movements of metal plate.
		}
		\label{fig:amp_cmp}
	\end{minipage}	
\end{figure*}

\textbf{Experimental Results.}
Fig.~\ref{fig:amp_cmp} shows the raw amplitude waveforms of two antennas respectively as well as the ratio of CSI amplitdue when the metal plate moves from $D_s$ = 5.83\,m to $D_e$ = 5.99\,m which incurs a reflection path length change of $2\sqrt{(\frac{5.5}{2})^{2}+5.99^{2}}-2\sqrt{(\frac{5.5}{2})^{2}+5.83^{2}}$ = 0.29\,m. According to the Fresnel zone model proposed in \cite{wang2016human, wu2016widir}, at the carrier frequency of 5.24\,GHz ($\lambda$ = 5.7\,cm), we ought to observe $\frac{0.29}{0.057}$ = 5 peaks/valleys in the CSI amplitude waveform.
Fig.~\ref{fig:amp_cmp} (a) and (b) show the CSI amplitudes retrieved from two antennas at the same receiver.
As shown in Fig.~\ref{fig:amp_cmp} (a), the plate movement-induced signal variation pattern is buried in the noise, and we can hardly see it.
The signal variation pattern is slightly clearer in Fig.~\ref{fig:amp_cmp} (b) but it is still difficult to be visualized. 
Fig.~\ref{fig:amp_cmp} (c) shows the ratio of amplitude between two antennas, and we can observe much clearer peaks/valleys. This is because the division operation cancels out most of the noise in the raw amplitude readings (e.g., high amplitude impulse and burst noise).
Thus, the signal variation pattern with amplitude ratio is much clearer than that with raw amplitude readings. When we vary the LoS path length, we have the same observation that the amplitude ratio shows much clearer movement-induced signal variation patterns especially when the target is far away from the transceivers.

The ratio of two CSI readings is still a complex number with amplitude and phase.
Mathematically, the amplitude is the ratio of the raw CSI amplitudes while the phase is the phase difference of the raw CSI phases.
We have just shown that the ratio of the amplitudes exhibits better sensing performance compared to raw amplitudes, we further explore the other properties of CSI ratio for human sensing.
\section{The CSI-Ratio Model}
\label{sec:model}
In this section, we first introduce the basic concept of CSI and present the relationship between human's movement and the change of CSI readings.
Next, we present the CSI-ratio model which establishes the corresponding relationship between human's movement and CSI ratio.
At last, we summarize three key properties of CSI ratio which can be utilized for human sensing and verify them via benchmark experiments.

\subsection{CSI Primer}
\label{sec:model:1}
In an indoor environment, radio frequency (RF) signals propagate from transmitter to receiver through multiple paths, i.e., one direct path and multiple reflection paths from objects (such as walls, furniture and the human target). The channel state information (CSI), which characterizes the multipath propagation, is a superposition of  signals from all the paths.
Mathematically, the CSI can be represented as:
\begin{equation}
\label{equation:csi1}
H(f,t) = \sum_{i=1}^L A_ie^{-j2{\pi}\frac{d_i(t)}{\lambda}}
\end{equation}
where $L$ is the number of paths, $A_i$ is the complex attenuation and $d_i(t)$ is the propagation length of the $i^{th}$ path. 

According to prior work \cite{wang2015understanding}, the paths can be grouped into static path and dynamic path.
Without loss of generality, we assume there is only one reflection path. When there is only one reflection path corresponding to the human target's movement, the dynamic component is the path reflected from the human target while the static component is composed of the LoS propagation and other reflection paths from static objects in the environment.
Thus, the CSI can be rewritten as:
\begin{equation}
\label{equation:csi2}
H(f,t)=H_s(f,t)+H_d(f,t)
=H_s(f,t)+A(f,t)e^{-j2\pi \frac{d(t)}{\lambda}}
\end{equation}
where $H_s(f,t)$ is the static component, $A(f,t)$, $e^{-j2\pi \frac{d(t)}{\lambda}}$ and $d(t)$ are the complex attenuation, phase shift and path length of dynamic component $H_d(f,t)$, respectively.

\begin{figure*}[t]
	\begin{minipage}[t]{0.27\linewidth}
		\centering
		\includegraphics[width=1\textwidth]{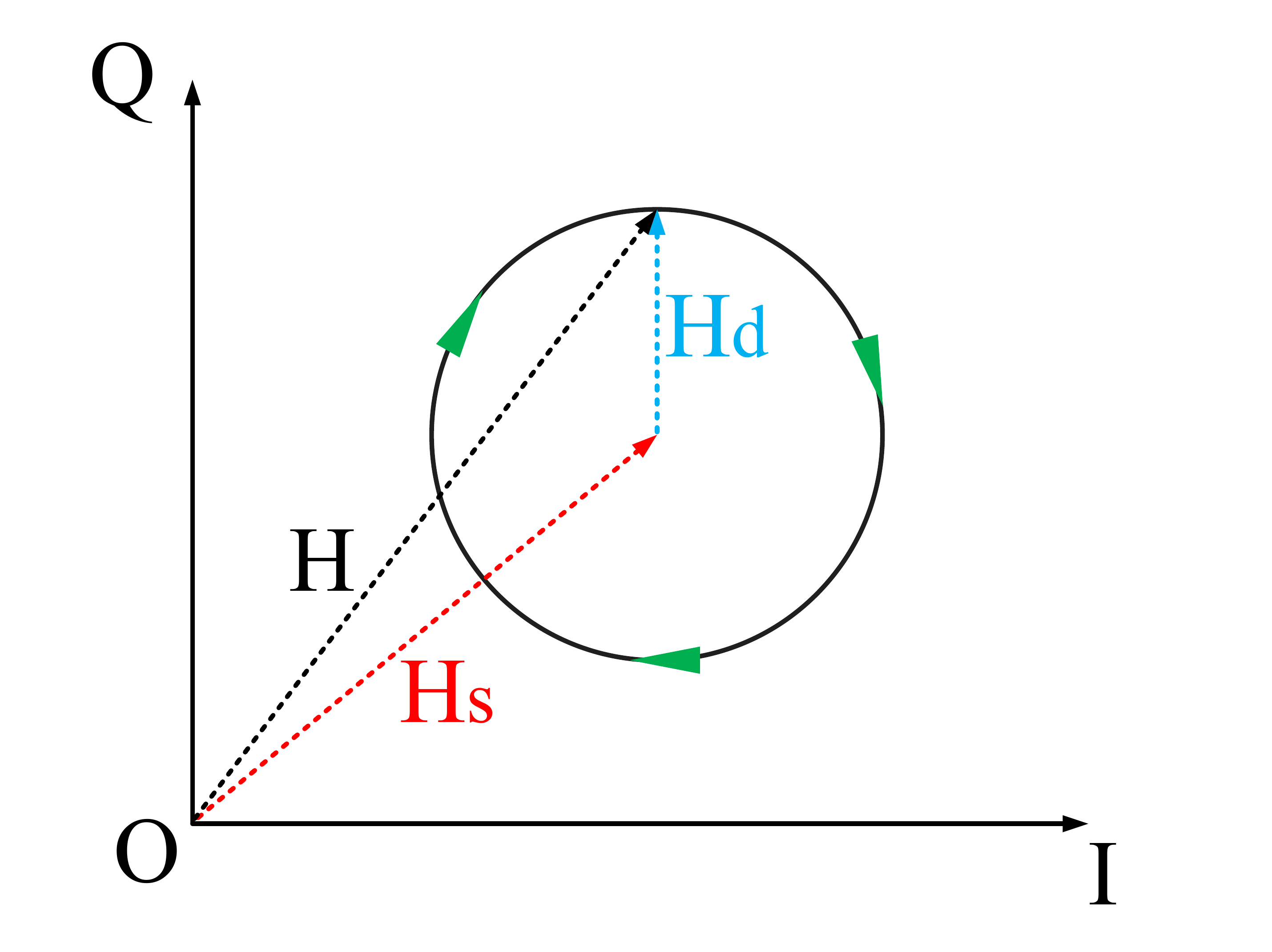}
		\caption{
			Ideal CSI in complex plane: when $d(t)$ increases, it rotates clockwise.
			In this case, its locus is a clockwise circle.
		}
		\label{fig:csi_phasor}
	\end{minipage}
	\hspace{2pt}
	\begin{minipage}[t]{0.71\linewidth}
		\centering
		\includegraphics[width=1\textwidth]{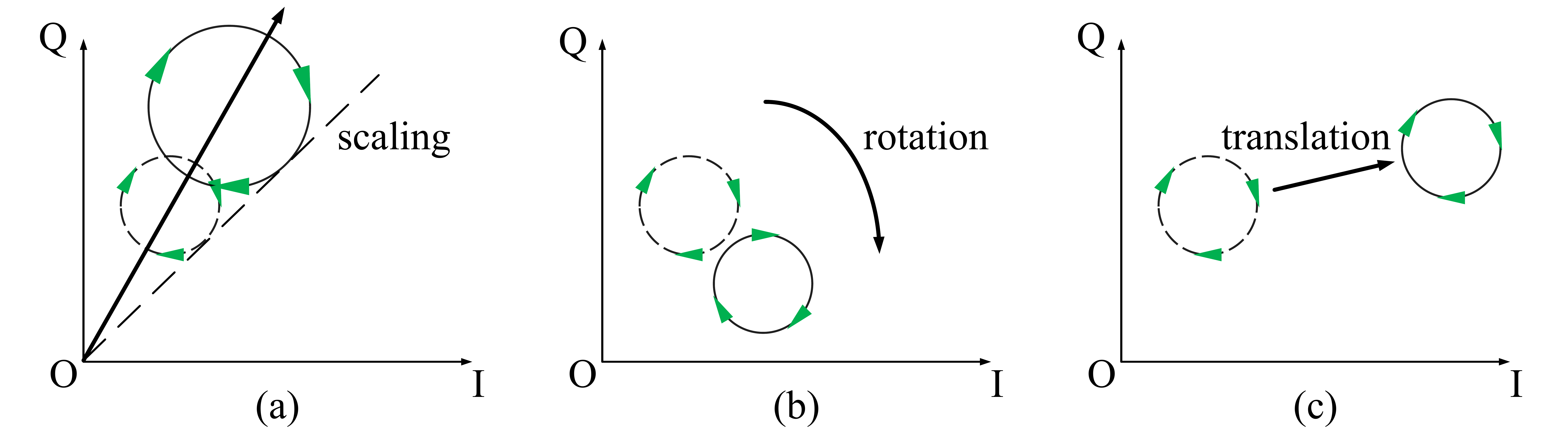}
		\caption{
			The scaling ($H(f,t) \mapsto \alpha H(f,t), \alpha \in \mathbb{R}$), rotation ($H(f,t) \mapsto e^{i\theta} H(f,t), \theta \in \mathbb{R}$) and translation ($H(f,t) \mapsto H(f,t)$ + $\beta, \beta \in \mathbb{C}$) operations do not change the geometric shape and rotation orientation (clockwise or counterclockwise) of the circle (represented as $H(f,t)$).
		}
		\label{fig:circle2}
	\end{minipage}	
\end{figure*}

When the human target moves a short distance, the signal amplitude of the dynamic component $A(f,t)$ can be considered as a constant. This is because the signal amplitude is determined by the path length. Changes of a few centimeters in path length have very little effect when the path length $d(t)$ is in the scale of meters.
When $d(t)$ is increased by one wavelength, the CSI ($H(f,t)$ in Eq.~\ref{equation:csi2}) rotates $2\pi$ clockwise, as shown in Fig.~\ref{fig:csi_phasor}.
It is easy to see that with operations such as scaling ($H(f,t) \mapsto \alpha H(f,t), \alpha \in \mathbb{R}$), rotation ($H(f,t) \mapsto e^{i\theta} H(f,t), \theta \in \mathbb{R}$) and translation ($H(f,t) \mapsto H(f,t)$ + $\beta, \beta \in \mathbb{C}$) in complex plane, the CSI trajectory is still a clockwise circle, as shown in Fig.~\ref{fig:circle2}.
This observation is quite important and helps to investigate the properties of CSI ratio in Sec.~\ref{sec:model:2}.
Note that for some sensing applications (e.g., respiration sensing and finger tracking), the caused path length change is smaller than one wavelength. In this case, $d(t)$ is increased by less than one wavelength and the CSI rotates less than $2\pi$ clockwise, thus its locus is just part of the full circle (a circular arc). 

Unfortunately, for commodity WiFi devices, as the transmitter and receiver are not time-synchronized, there is a time-varying random phase offset $e^{-j\theta_{offset}}$ in each CSI sample as follows:
\begin{equation}
\label{equation:csi3}
H(f,t)=e^{-j\theta_{offset}} (H_s(f,t)+A(f,t)e^{-j2\pi \frac{d(t)}{\lambda}})
\end{equation}
With this random phase offset, the movement-induced CSI change in complex plane is no longer a circle. This random phase offset thus prevents us from directly using the CSI phase information for fine-grained sensing.

\subsection{Understanding CSI Ratio}
\label{sec:model:2}
Before derivating the formula of CSI ratio, we first present two key observations as follows:
\begin{enumerate}
	\item For commodity WiFi card such as the widely used Intel 5300, the time-varying phase offset is the same across different antennas on a Wi-Fi card as they share the same RF oscillator~\cite{kotaru2015spotfi, li2016dynamic}. 
	\item When the target moves a short distance (a few centimeters), the difference of the two reflection path lengths at two close-by antennas $d_2(t)-d_1(t)$ can be considered as a constant $\Delta d$ \cite{zeng2018fullbreathe}.
\end{enumerate}

Based on these two observations, we obtain the CSI ratio by taking the ratio of CSI in Eq.~\ref{equation:csi3} between two antennas as follows:
\begin{equation}
\label{equation:csiq}
\begin{split}
	\dfrac{H_1(f,t)}{H_2(f,t)} 
	&=\dfrac{e^{-j\theta_{offset}} (H_{s,1}+A_1e^{-j2\pi \frac{d_1(t)}{\lambda}})}{e^{-j\theta_{offset}} (H_{s,2}+A_2e^{-j2\pi \frac{d_2(t)}{\lambda}})}\\
	&=\dfrac{A_1e^{-j2\pi \frac{d_1(t)}{\lambda}}+H_{s,1}}{A_2e^{-j2\pi \frac{d_1(t)+\Delta d}{\lambda}}+H_{s,2}}\\
	&=\dfrac{A_1e^{-j2\pi \frac{d_1(t)}{\lambda}}+H_{s,1}}{A_2e^{-j2\pi \frac{\Delta d}{\lambda}}e^{-j2\pi \frac{d_1(t)}{\lambda}}+H_{s,2}}
\end{split}
\end{equation}
where $H_{1}(f,t)$ is the CSI of the first antenna and $H_{2}(f,t)$ is the CSI of the second antenna.
To simplify the equation for easier illustration, we employ $\mathcal{A}, \mathcal{B}, \mathcal{C}, \mathcal{D}$ and $\mathcal{Z}$ to represent the terms: $A_1=\mathcal{A}$, $H_{s,1}=\mathcal{B}$, $A_2e^{-j2\pi \frac{\Delta d}{\lambda}}=\mathcal{C}$ and $H_{s,2}=\mathcal{D}$; $e^{-j2\pi \frac{d_1(t)}{\lambda}}=\mathcal{Z}$ represents a unit circle rotates clockwise when $d_1(t)$ increases.
We can then simplify Eq.~\ref{equation:csiq} as:
\begin{equation}
\label{equation:mobius}
	\dfrac{H_1(f,t)}{H_2(f,t)} =\dfrac{\mathcal{AZ}+\mathcal{B}}{\mathcal{CZ}+\mathcal{D}}
\end{equation}
which is exactly in the form of Mobius transformation \cite{schwerdtfeger1979geometry}, provided $\mathcal{BC-AD} \neq 0$.
We further decompose it into the following form:
\begin{equation}
\label{equation:mobius_decomposed}
	\dfrac{H_1(f,t)}{H_2(f,t)} =\dfrac{\mathcal{BC}-\mathcal{AD}}{\mathcal{C}^2} \cdot \dfrac{1}{\mathcal{Z}+\frac{\mathcal{D}}{\mathcal{C}}} + \dfrac{\mathcal{A}}{\mathcal{C}}
\end{equation}
Consequently, the mapping in Eq.~\ref{equation:mobius_decomposed} is composed of the following transformations \cite{needham1998visual}:
\begin{equation}
\label{equation:four_trans}
	\left.
	\begin{aligned}
		&\text{(\romannumeral 1)} \quad\ \ \mathcal{Z} \mapsto \mathcal{Z}+\frac{\mathcal{D}}{\mathcal{C}} \text{, which is a translation by}\ \frac{\mathcal{D}}{\mathcal{C}}\text{;}\\
		&\text{(\romannumeral 2)} \quad\ \mathcal{Z} \mapsto \frac{1}{\mathcal{Z}} \text{, which is a complex inversion;}\\ 
		&\text{(\romannumeral 3)} \quad \mathcal{Z} \mapsto \frac{\mathcal{BC}-\mathcal{AD}}{\mathcal{C}^2}\mathcal{Z} \text{, which is a multiplication by the complex number $\frac{\mathcal{BC}-\mathcal{AD}}{\mathcal{C}^2}$;\footnotemark[2]}\\
		&\text{(\romannumeral 4)} \quad \mathcal{Z} \mapsto \mathcal{Z}+\frac{\mathcal{A}}{\mathcal{C}} \text{, which is another translation by}\ \frac{\mathcal{A}}{\mathcal{C}}\text{.}
	\end{aligned}
	\right\}
\end{equation}
\footnotetext[2]{Mathematically, a multiplication by a complex number $Ae^{i\theta}$ corresponds to a scaling (amplitude $A$) and a rotation (phase $\theta$).}

\begin{figure*}[t]
	\begin{minipage}[t]{0.96\linewidth}
		\centering
		\includegraphics[width=0.96\textwidth]{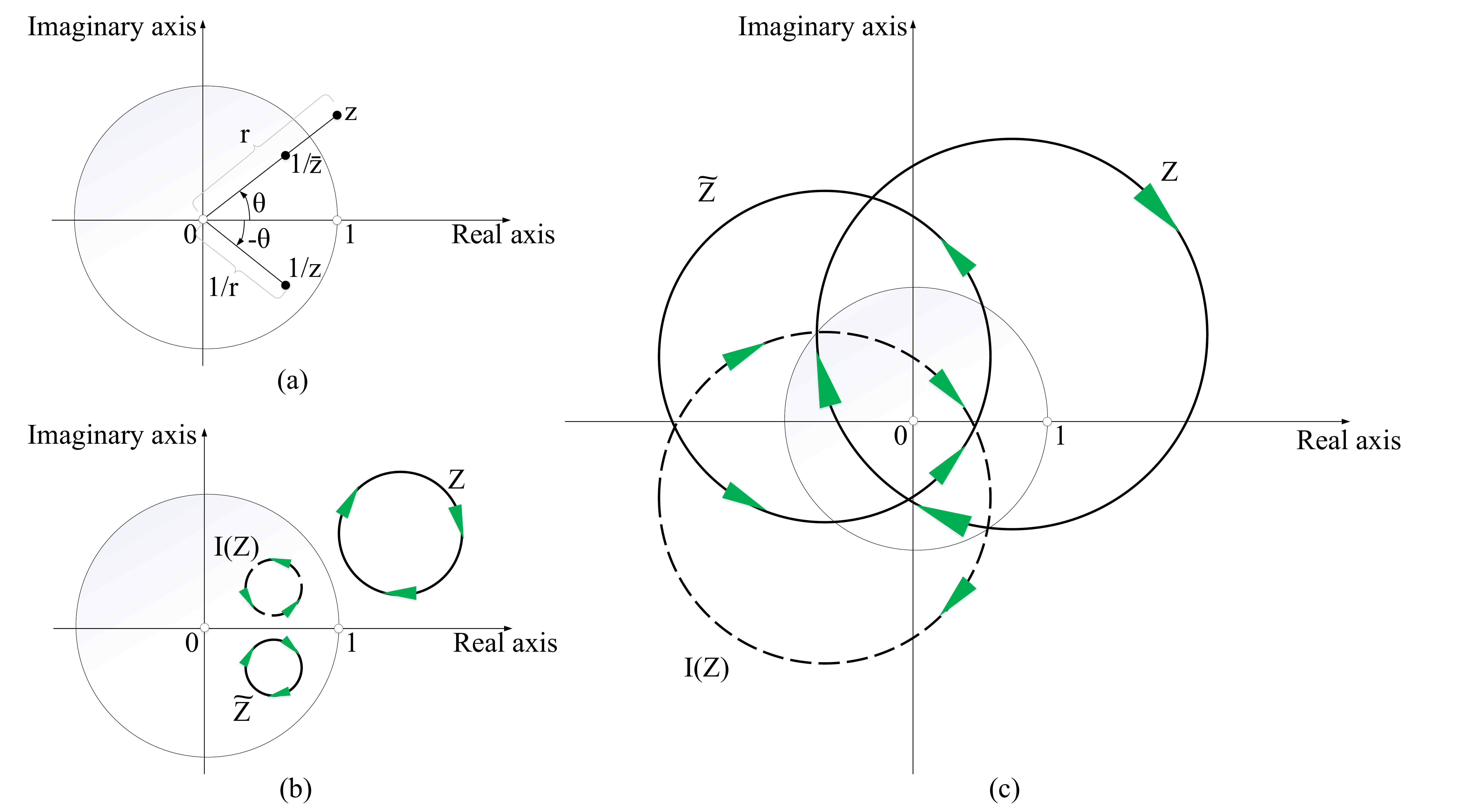}
		\caption{Illustration of complex inversion: (a) the point $z$ is mapped to a new point $\frac{1}{z}$ that has the reciprocal length and the negative angle; (b) the clockwise circle $\mathcal{Z}$ that does not contain the origin is mapped to a new circle $\widetilde{\mathcal{Z}}$ that has the same orientation; (c) the clockwise circle $\mathcal{Z}$ that contains the origin is mapped to a new circle that has the opposite orientation. Here, $1/\overline{z}$ and $I(\mathcal{Z})$ represent the inversion of the point $z$ and the circle $\mathcal{Z}$, respectively.}
		\label{fig:complex_inversion}
	\end{minipage}
\end{figure*}

As shown in Sec.~\ref{sec:model:1}, the scaling and rotation operation in step (\romannumeral 3) and translation operations in step (\romannumeral 1) and (\romannumeral 4) do not change the geometric shape and rotation orientation (clockwise or counterclockwise) of the circle.
Thus, for the remaining step (\romannumeral 3) in Eq.~\ref{equation:four_trans}, its complex inversion operation ($\mathcal{Z} \mapsto \frac{1}{\mathcal{Z}}$) holds the key to understanding the effect of CSI ratio (Mobius transformation) and we introduce it below.

Mathematically, the complex inversion of a point $z=re^{i\theta}$ is $\frac{1}{r}e^{-i\theta}$: the new length is the reciprocal of the original, and the new phase is the opposite of the original.
Fig.~\ref{fig:complex_inversion} (a) shows how a point $z$ outside the unit circle is mapped to a point $\frac{1}{z}$ inside it: (1) move $z=re^{i\theta}$ to the point that is at the same direction as $z$ but has a reciprocal length, namely the point $\frac{1}{r} e^{i\theta}$;
(2) apply a complex conjugate operation~(i.e., symmetrical with respect to the real axis) to arrive at the point $\frac{1}{r} e^{-i\theta}$.

Next, we investigate the effect of complex inversion (denoted as $\widetilde{\mathcal{Z}}$) on a clockwise circle $\mathcal{Z}$, as shown in Fig.~\ref{fig:complex_inversion} (b) and (c). The complex inversion does not change the shape of $\mathcal{Z}$ \cite{needham1998visual}.
However, it may change the rotation orientation of $\mathcal{Z}$.  Whether $\widetilde{\mathcal{Z}}$ has the same orientation as $\mathcal{Z}$ depends on whether the origin (0, 0) is inside $\mathcal{Z}$ or not.
As shown in Fig.~\ref{fig:complex_inversion} (b), if (0, 0) is not inside $\mathcal{Z}$, then $\widetilde{\mathcal{Z}}$ has the same rotation orientation as $\mathcal{Z}$.
As shown in Fig.~\ref{fig:complex_inversion} (c), if (0, 0) is inside $\mathcal{Z}$, then $\widetilde{\mathcal{Z}}$ has the opposite rotation orientation.
Recall that the complex inversion is applied to $\mathcal{Z}+\frac{\mathcal{D}}{\mathcal{C}}$ in Eq.~\ref{equation:mobius_decomposed} where $\mathcal{Z}$ represents a unit circle and $\frac{\mathcal{D}}{\mathcal{C}}$ is the translation to the origin (0, 0).
Thus, given $|\frac{\mathcal{D}}{\mathcal{C}}|=|\frac{H_{s,2}}{A_2}|>1$ (the radius of a unit circle), $\mathcal{Z}+\frac{\mathcal{D}}{\mathcal{C}}$ will not contain the origin.
So we can conclude that $\frac{1}{\mathcal{Z}+\frac{\mathcal{D}}{\mathcal{C}}}$ does have the same orientation as $\mathcal{Z}+\frac{\mathcal{D}}{\mathcal{C}}$.
As shown above, the scaling, rotation and translation in Eq.~\ref{equation:mobius_decomposed} does not change the rotation orientation, thus the CSI ratio $\frac{\mathcal{AZ}+\mathcal{B}}{\mathcal{CZ}+\mathcal{D}}$ has the same rotation orientation as $\frac{1}{\mathcal{Z}+\frac{\mathcal{D}}{\mathcal{C}}}$ (and $\mathcal{Z}+\frac{\mathcal{D}}{\mathcal{C}}$).
This is also the case for human sensing most of the time: the magnitude of the static component $|H_{s,2}|$ is larger than that of the dynamic component $|A_{2}|$ \cite{li2017indotrack}.
This may not hold true when the LoS path signal is attenuated by objects, in which case the CSI ratio $\frac{\mathcal{AZ}+\mathcal{B}}{\mathcal{CZ}+\mathcal{D}}$ has the opposite rotation orientation as $\mathcal{Z}+\frac{\mathcal{D}}{\mathcal{C}}$.

Based on the analysis above, it is easy to obtain the following three properties of CSI ratio for human sensing:
\begin{enumerate}
	\item[P1] The CSI ratio changes following a circle in complex plane when the reflection path length changes several wavelengths.
	\item[P2] If the reflection path length increases: when the magnitude of static component is larger than that of dynamic component, CSI ratio rotates clockwise; otherwise, it rotates counterclockwise.
	\item[P3] If the reflection path length changes exactly one wavelength, the CSI ratio forms a full circle whose radian is exactly $2\pi$ in complex plane. If the reflection path length changes less than one wavelength, the CSI ratio forms a circular arc whose radian roughly matches the reflection path length change.
\end{enumerate}
These three properties in CSI-ratio model relate the reflection path length change to the change of CSI ratio.

\subsection{Model Verification with Benchmark Experiments}
\label{sec:model:3}
\begin{figure*}[t]
	\begin{minipage}[t]{0.41\linewidth}
		\centering
		\includegraphics[width=1\textwidth]{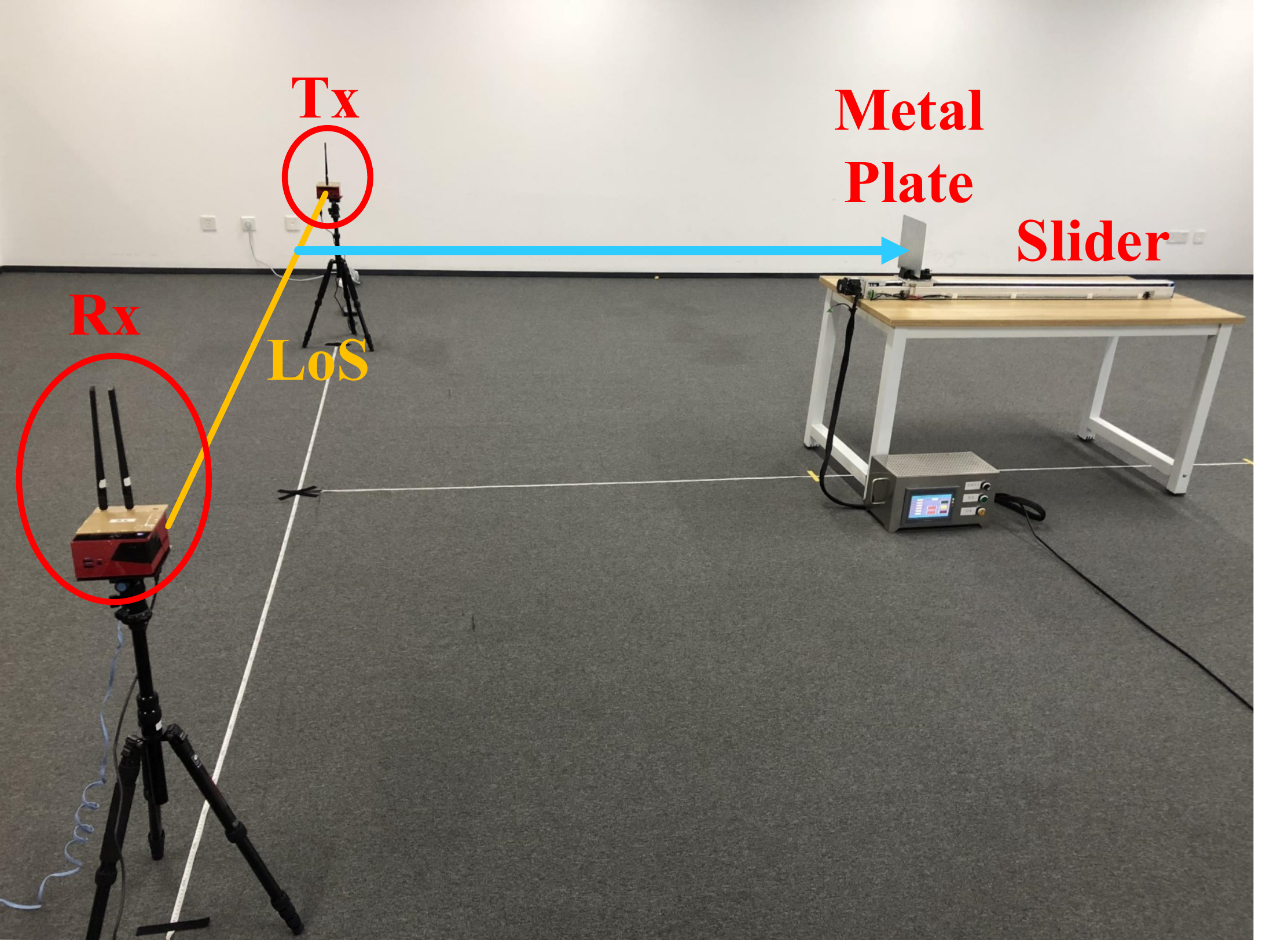}
		\caption{Experimental settings for model verification.}
		\label{fig:model_setting}
	\end{minipage}
	\hspace{2pt}
	\begin{minipage}[t]{0.56\linewidth}
		\centering
		\includegraphics[width=1\textwidth]{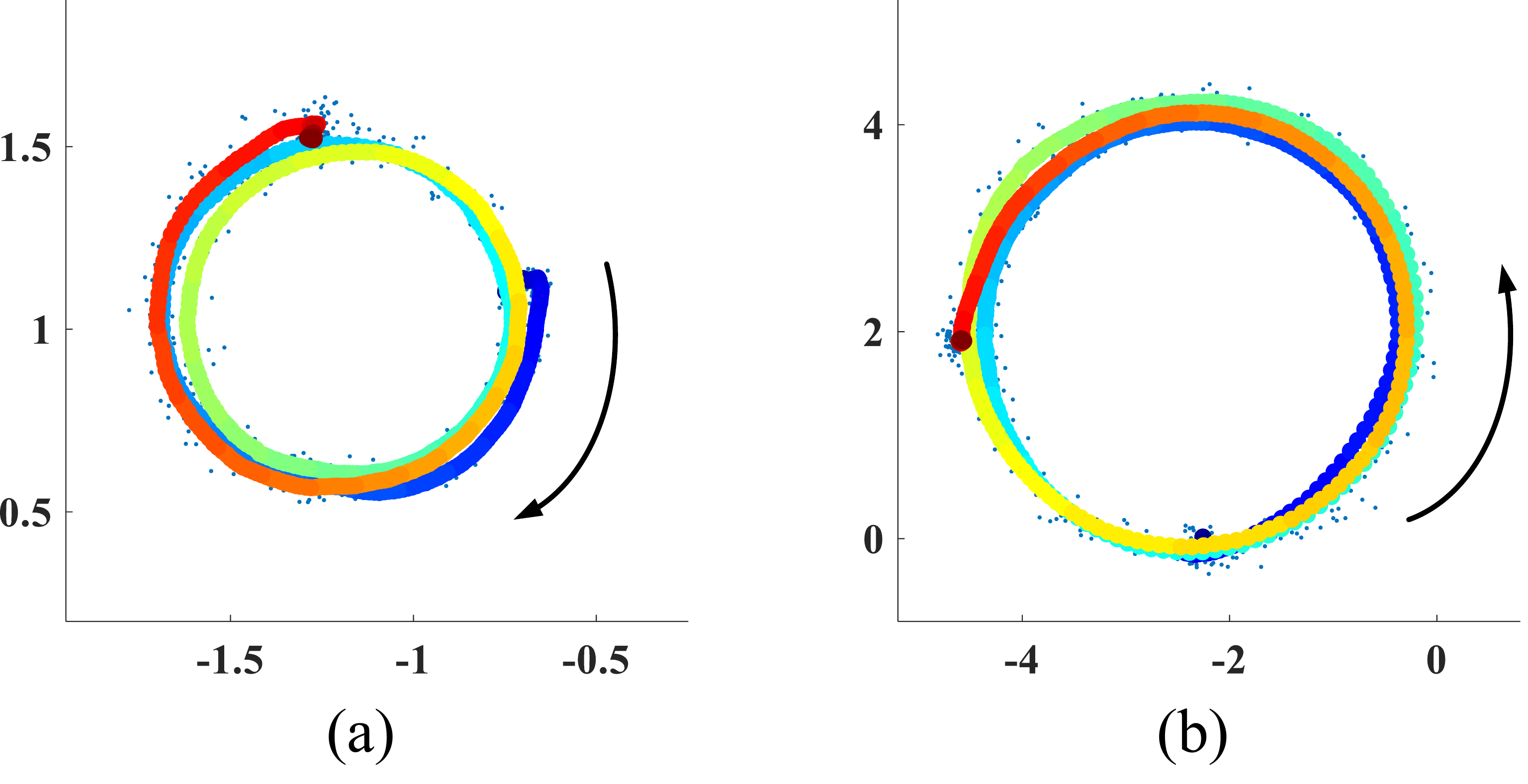}
		\caption{The CSI ratio in complex plane when the metal plate moves 0.1\,m away from the LoS path: (a) normal case; (b) the LoS path is obscured by a large metal plate between the transceivers.}
		\label{fig:exp_results1}
	\end{minipage}
\end{figure*}

In this subsection, we verify the three properties above via benchmark experiments.
As shown in Fig.~\ref{fig:model_setting} (a), the verification experiments are conducted in a large room with less multipath so that the three properties can be more clearly visualized.   
We setup one pair of GIGABYTES mini-PCs equipped with Intel 5300 NICs as transceivers, and the LoS distance is set as 4\ meters.
The transmitter (Tx) is equipped with one antenna while the receiver (Rx) is equipped with two antennas. All the antennas are commonly-seen vertically-polarized omni-directional antennas. The central frequency is set as 5.24\,GHz, corresponding to the wavelength of 5.725\,cm.
We adopt a metal plate as the reflector to reflect off WiFi signals. To precisely control the displacement of the metal plate, we mount it on a high-precision linear motion slider with an accuracy of 0.01\,mm.
We put a desk at the perpendicular bisector of the LoS path and place the slider on the desk, as shown in Fig.~\ref{fig:model_setting}.
When the metal plate moves along the slider, the reflection path length changes accordingly.
Note that, in this real-life setting, the magnitude of static component is much larger than that of dynamic component.

\textbf{Verification of P1.}
We control the metal plate to move 0.1\,m away from the LoS path with the start position of 2.55\,m perpendicular to the LoS.
As shown in Fig.~\ref{fig:exp_results1} (a), the dots are the samples of CSI ratio and the line is smoothed CSI ratio after Savitzky-Golay filter which can effectively preserve the envelope of the raw waveform \cite{baba2014enhancing}.
We can observe that the CSI ratio changes following a circle in complex plane, which meets our expectation.
We obtain the same results when the plate moves at different start positions.

\textbf{Verification of P2.}
When the LoS path is not obscured, the magnitude of static component is larger than that of dynamic component, in which case the CSI ratio rotates clockwise, as shown in Fig.~\ref{fig:exp_results1} (a).
Next, we place a large square metal plate (1\,m $\times$ 1\,m) vertically between the transceivers to greatly attenuate the LoS path signal so that the magnitude of the dynamic component is larger than that of the static component.
Then we repeat the experiment as the same as verification of P1.
In this case, when the metal plate moves 0.1\,m away from the LoS path, the CSI ratio rotates counterclockwise, as shown in Fig.~\ref{fig:exp_results1} (b).
We also conduct the experiments when the metal plate moves towards the LoS path and observe that if the reflection path length decreases: when the magnitude of static component is larger than that of dynamic component, CSI ratio rotates counterclockwise; otherwise, it rotates clockwise.
We skip these figures for conciseness.

\begin{figure*}[t]
	\begin{minipage}[t]{0.43\linewidth}
		\centering
		\includegraphics[width=1\textwidth]{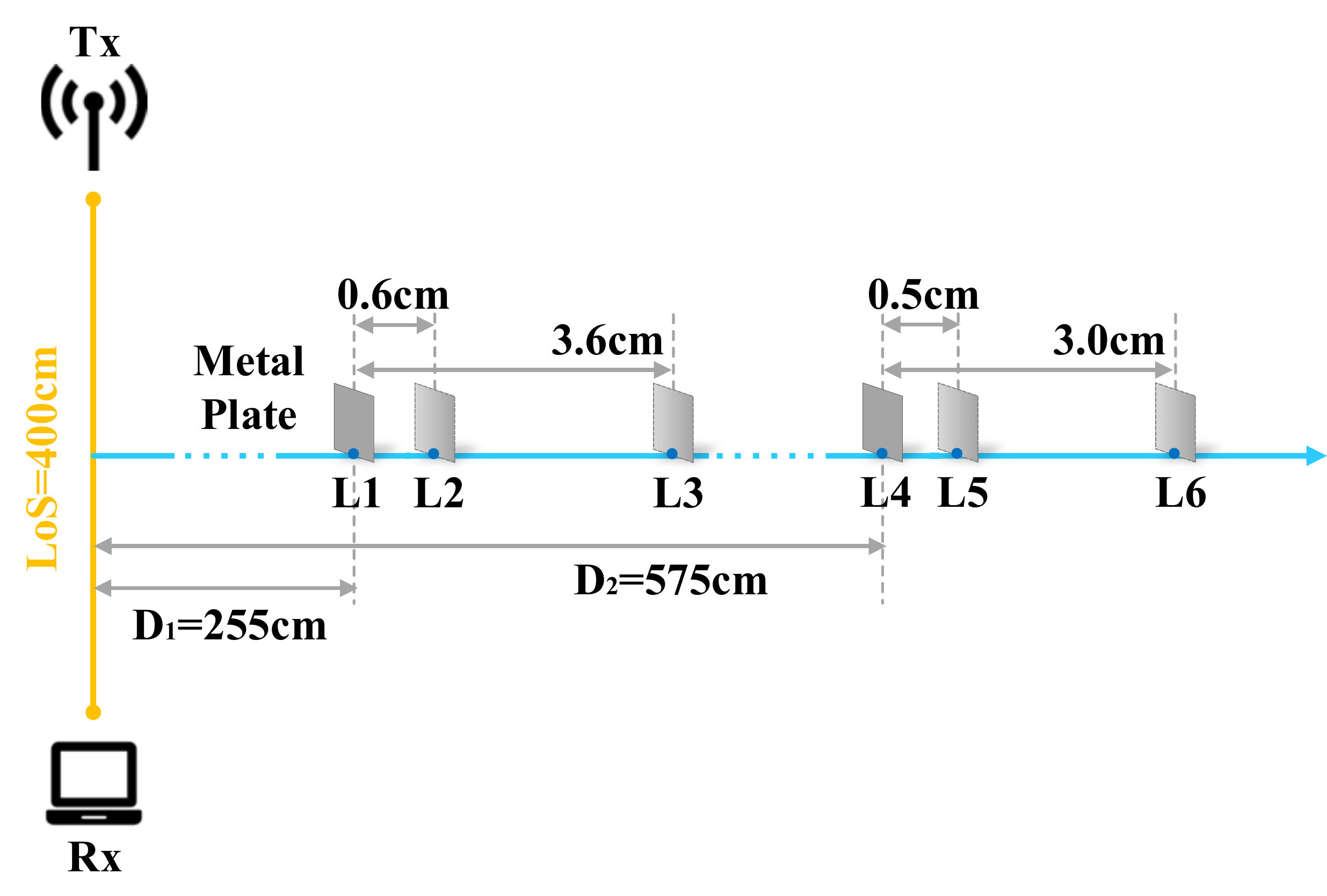}
		\caption{Conceptual illustration of experimental settings for verification of P3.
		}
		\label{fig:conceptual}
	\end{minipage}
	\hspace{2pt}
	\begin{minipage}[t]{0.54\linewidth}
		\centering
		\includegraphics[width=1\textwidth]{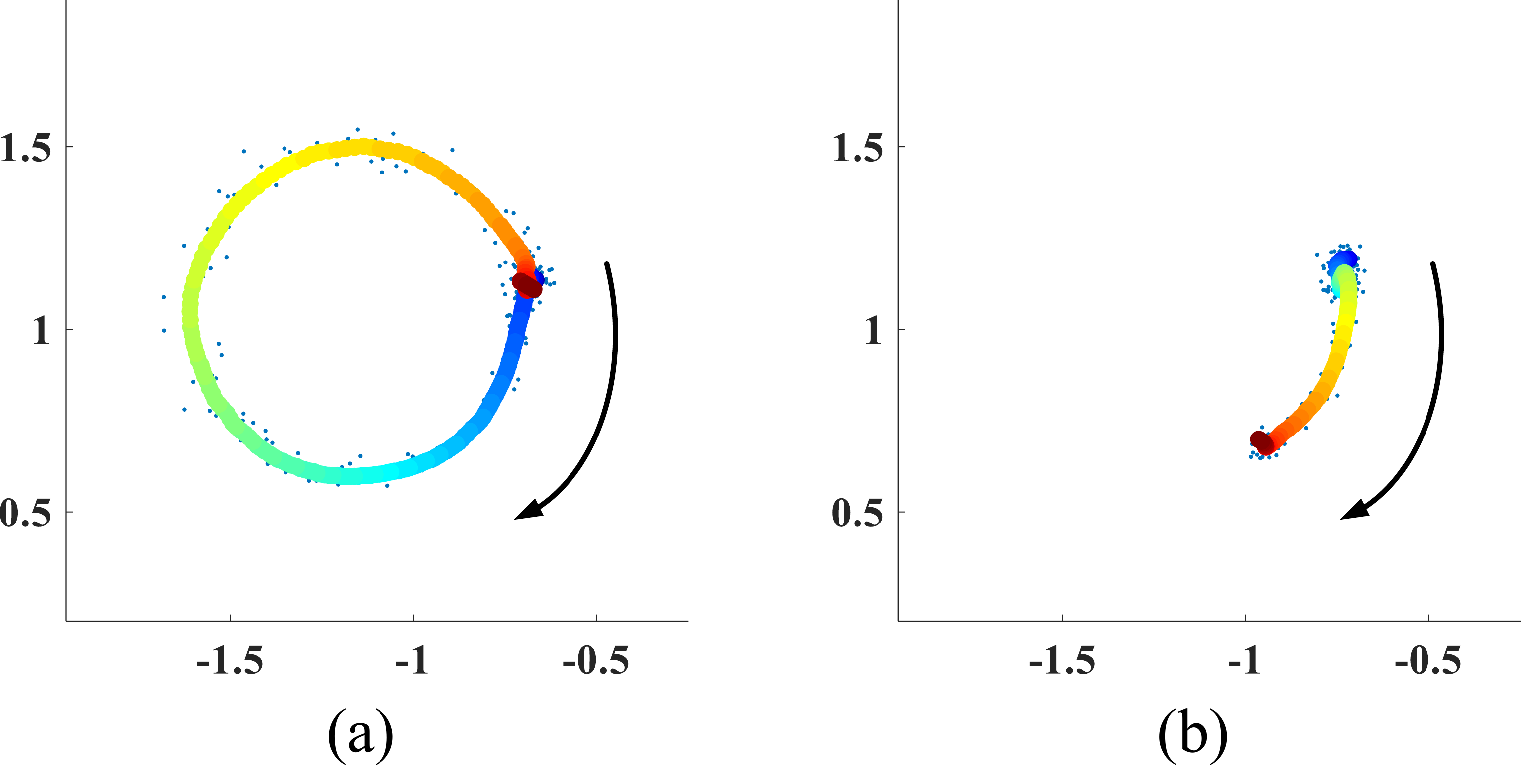}
		\caption{The CSI ratio in complex plane when the metal plate moves away from the LoS path: (a) from \emph{L1} to \emph{L3}; (b) from \emph{L1} to \emph{L2}.
		}
		\label{fig:exp_results2}
	\end{minipage}
\end{figure*}

\textbf{Verification of P3.}
As shown in Fig.~\ref{fig:conceptual}, we conduct experiments in two different regions, one near to the LoS (255\,cm) and the other relatively far away (575\,cm).
In each region, we control the start and end positions of the metal plate's motion to make sure the reflection path length changes one wavelength or one-sixth of the wavelength.
In detail, when the metal plate moves from \emph{L1} (255\,cm) to \emph{L3} (258.6\,cm) and from \emph{L4} (575\,cm) to \emph{L6} (578.0\,cm), the reflection path length changes exactly one wavelength.
When the metal plate moves from \emph{L1} (255\,cm) to \emph{L2} (255.6\,cm) and from \emph{L4} (575\,cm) to \emph{L5} (575.5\,cm), the reflection path length changes exactly one-sixth of the wavelength.
Note that the experiments described above are conducted without obscuration of the LoS path, which means the magnitude of static component is larger than that of dynamic component.
Fig.~\ref{fig:exp_results2} (a) shows the CSI ratio when the metal plate moves from \emph{L1} to \emph{L3}, and we can observe a clear full circle.
That is to say, when the metal plate moves from \emph{L1} to \emph{L3}, the reflection path length is changed by one wavelength, and thus the CSI ratio rotates clockwise by exactly $2\pi$ along a circle.
In Fig.~\ref{fig:exp_results2} (b), we observe the CSI ratio rotates clockwise by roughly $\frac{\pi}{3}$ along a circle when the metal plate moves from \emph{L1} to \emph{L2}.
That is to say, when the reflection path length changes less than one wavelength, the CSI ratio forms a circular arc whose radian roughly matches the reflection path length change.
We obtain similar results when the plate is 5.75 meters away, and we omit the figures for the sake of brevity.

To sum up, the benchmark results above validate the three properties of CSI ratio.
These three properties relate the reflection path length change to the change of CSI ratio and guide us how to sense human activities exploiting CSI ratio.
We want to point out that the proposed CSI-ratio model in this session is a very general method which can be applied to many existing WiFi-based sensing applications such as indoor tracking and motion detection.
It also lays a solid foundation for sensing finer-grained subtle movement, such as tiny finger tracking.
It can also be applied to other wireless technologies such as RFID, LTE and LoRa.
In this paper, we focus on applying CSI ratio to significantly increase the sensing range of WiFi-based respiration, and we present the details in next section.	
\section{Extracting Respiration Pattern from CSI Ratio}
\label{sec:pattern_extraction}

In this section, we aim to extract the subtle signal variation caused by respiration from CSI ratio even when the target is far away.
We first introduce how to apply the CSI-ratio model presented in previous section for human respiration sensing.
Next, we present our approach that extracts respiration pattern more accurately by combining the amplitude and phase of the CSI ratio to achieve an even further sensing range and higher accuracy. 

\subsection{Applying CSI-ratio Model for Respiration Sensing}
\label{sec:pattern_extraction:1}

\begin{figure*}[t]
	\begin{minipage}[t]{0.88\linewidth}
		\centering
		\includegraphics[width=1\textwidth]{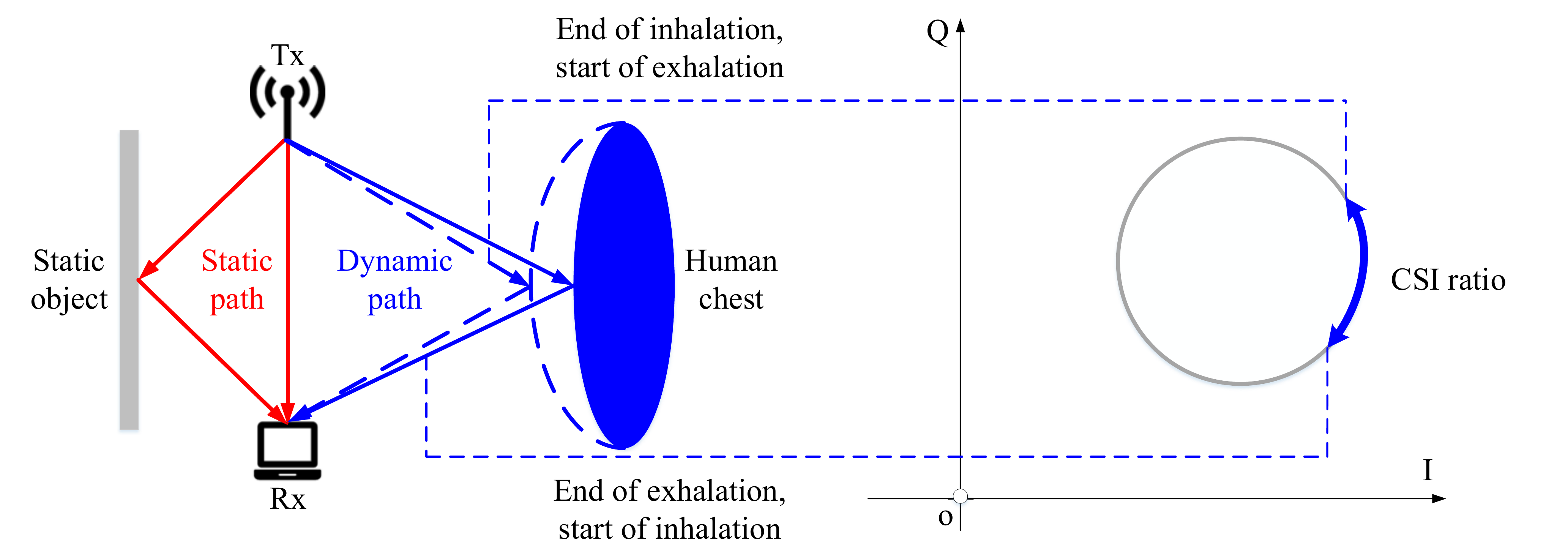}
		\caption{When a subject breathes, the dynamic path length increases or decreases, then CSI ratio rotates along a circular arc (labeled in blue) clockwise or counterclockwise corresponding to the inhalation and exhalation.}
		\label{fig:apply_csi_ratio}
	\end{minipage}		
\end{figure*}

For respiration sensing, it is safe to assume that there is one dominating reflection path from the human chest, as previously verified in~\cite{wang2016human, zeng2018fullbreathe}.
Therefore, the formula of CSI ratio obtained in Eq.~\ref{equation:csiq} can be directly applied for respiration sensing where $d(t)$ is the path length of signal bouncing off the human chest.
Since the chest displacement caused by respiration is between 5\,mm to 12\,mm \cite{lowanichkiattikul2016impact}, the reflection path length changes less than one wavelength (5.7\,cm for 5.24\,GHz), then the locus of CSI ratio during respiration is just a circular arc (part of a full circle).
As shown in Fig.~\ref{fig:apply_csi_ratio}, when a subject inhales and exhales during respiration, the dynamic reflection path length increases and decreases accordingly.  The CSI ratio rotates along the circular arc clockwise or counterclockwise corresponding to inhalation and exhalation, respectively. 
By analyzing the time interval between clockwise and counterclockwise rotation, we can obtain the respiration rate.

As shown in Sec.~\ref{sec:model:2}, when reflection path length increases, whether CSI ratio rotates clockwise or counterclockwise is determined by the relationship between the static and dynamic components.
However, the rotation orientation does not change the time interval between exhalation and inhalation during respiration and thus has no effect on the respiration rate estimation.

\subsection{Combining Amplitude and Phase of CSI Ratio}
\label{sec:pattern_extraction:2}
With the CSI reading at one antenna, the phase can not be utilized for sensing because of the time-varying random phase offset.
With CSI ratio, the phase difference between two antennas is stable (the random phase offset gets canceled out). Thus, we can combine the phase and amplitude of the CSI ratio to eliminate the "blind spots" issue reported in \cite{wang2016human} and further extend the sensing range.

Note that a complex number can be represented in the form of $a+bi$ as well as $Ae^{i\theta}$ where $a, b$ are the real part (I) and imaginary part (Q), and $A, \theta$ are the amplitude and phase, respectively.
The orthogonal I and Q components of CSI ratio keep perfect complementarity for respiration sensing which means at a location I component is bad for sensing, Q component is good and vice versa.

Thus, we combine the I, Q components for sensing which is equivalent to combining amplitude and phase. As shown below, the combination scheme consists of two steps: (1) generating multiple combination candidates by assigning different weights to the I/Q components for combination; (2) selecting one from the candidates as the final extracted respiration pattern.

\subsubsection{Generating Combination Candidates}
\label{sec:pattern_extraction:2:1}

\begin{figure*}[t]
	\begin{minipage}[t]{0.34\linewidth}
		\centering
		\includegraphics[width=0.96\textwidth]{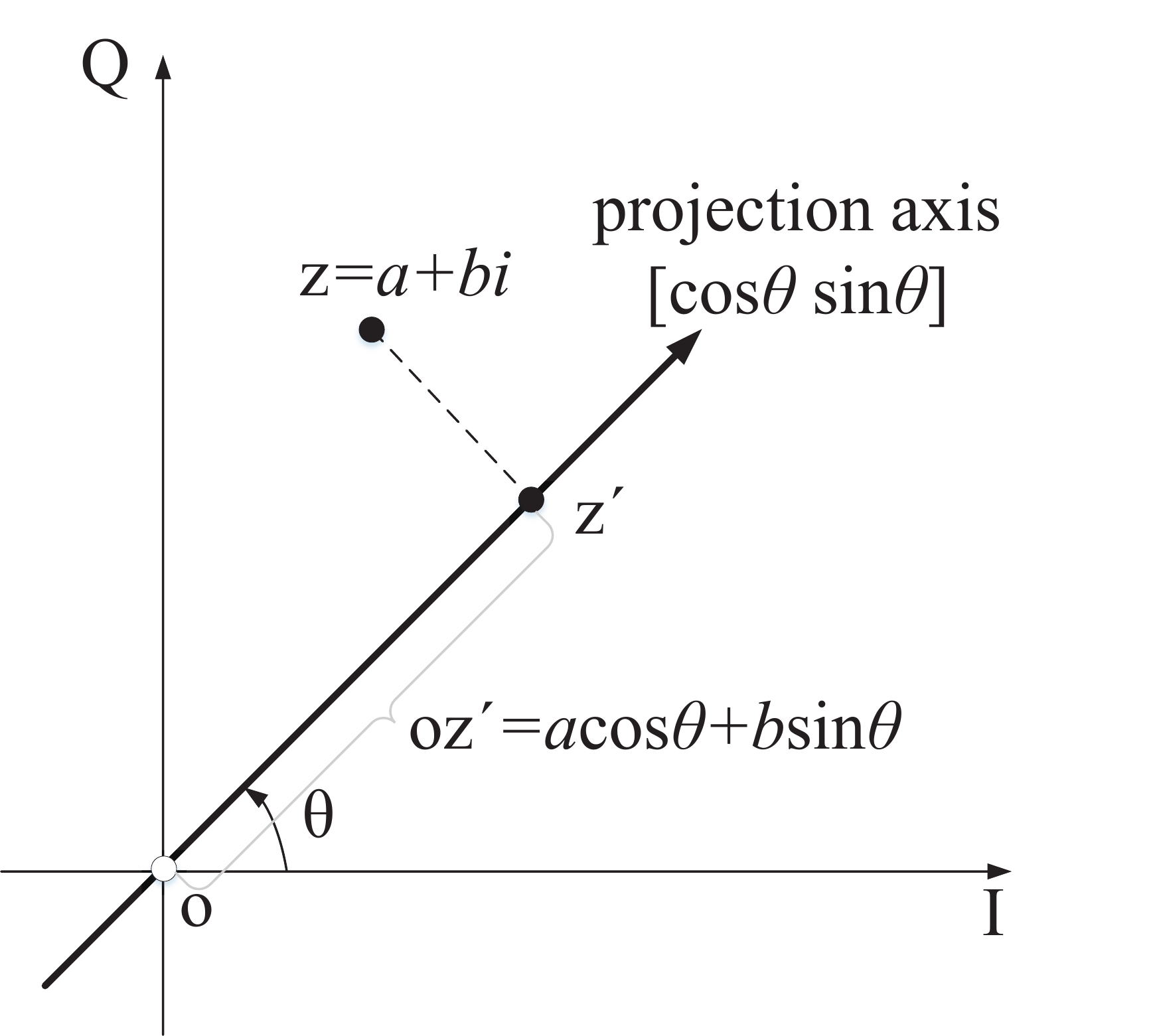}
		\caption{A point $z=a+bi$ is projected on an axis to get a new point $z'$.}
		\label{fig:IQ_combine}
	\end{minipage}
	\hspace{2pt}
	\begin{minipage}[t]{0.64\linewidth}
		\centering
		\includegraphics[width=1\textwidth]{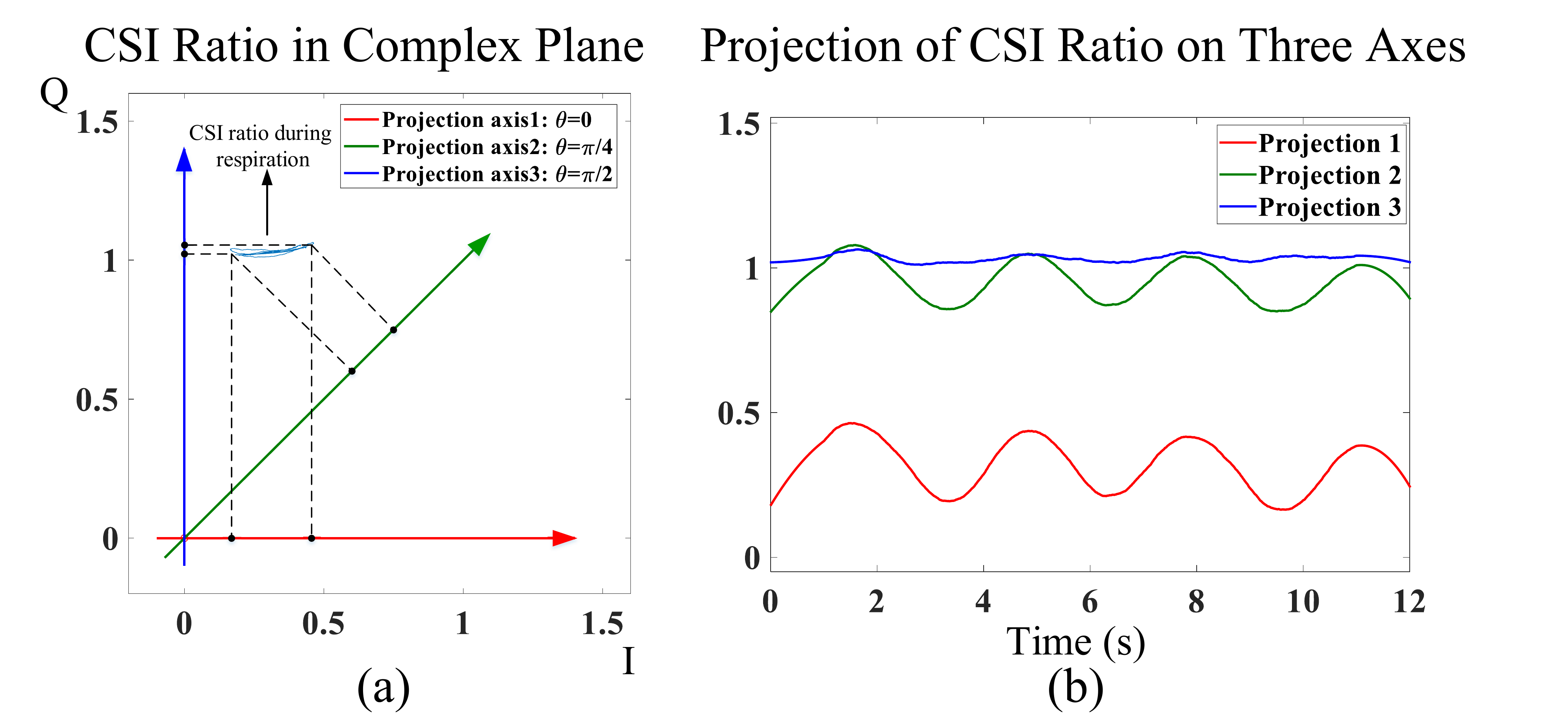}
		\caption{Example of projecting a time series of two-dimensional CSI ratio during respiration in complex plane: (a) CSI ratio and three projection axes ($\theta=0, \pi/4, \pi/2$); (b) projection results on three corresponding axes in (a).
		}
		\label{fig:projection_rationale}
	\end{minipage}		
\end{figure*}

Traditional approach~\cite{droitcour2004range} simply selects a better one from I and Q, thus has only two candidates--I and Q.
We linearly combine I/Q components by projecting the complex-valued CSI ratio on an axis in the complex plane.
Fig.~\ref{fig:IQ_combine} shows how a point $z=a+bi$ is projected on an axis $\begin{bmatrix}\cos\theta ~\sin\theta\end{bmatrix}$ to get a new point $z'$, where $\theta$ is the angle of projection axis.
Following simple geometry, we can obtain:
\begin{equation}
\begin{split}
	oz'&=\begin{bmatrix}\cos\theta~\sin\theta\end{bmatrix} \begin{bmatrix}a~b\end{bmatrix}^T \\
	&=a\cos\theta+b\sin\theta
\end{split}
\end{equation} 
which is exactly the linear combination of the I-Q components of the point $z$.
In this combination, the weights assigned are $\cos\theta$ for I component and $\sin\theta$ for Q component. Similarly, for a time series of CSI ratio data $\mathbf{x}$, its projection $\mathbf{y}$ on axis $\begin{bmatrix}\cos\theta~\sin\theta\end{bmatrix}$ can be denoted as:
\begin{equation}
\label{equation:pca}
\mathbf{y} = \begin{bmatrix}\cos\theta~ \sin\theta\end{bmatrix} \begin{bmatrix}\Re(\mathbf{x})~ \Im(\mathbf{x})\end{bmatrix}^T
\end{equation}
where $\Re(\mathbf{x})$ is the real part (I component) of $\mathbf{x}$ and $\Im(\mathbf{x})$ is the imaginary part (Q component) of $\mathbf{x}$.
By varying $\theta$ from 0 to $2\pi$ at a fixed step size, we can generate different combination candidates. 

Fig.~\ref{fig:projection_rationale} presents the example of projecting a time series of two-dimensional CSI ratio (during respiration) on three axes in complex plane.
As shown in Fig.~\ref{fig:projection_rationale} (a), the locus of CSI ratio during respiration process is a circular arc, and three different projection axes are chosen, namely axis~1~($\theta=0$), axis~2~($\theta=\pi/4$) and axis~3~($\theta=\pi/2$).
Projection axis~1 is tangential to the circular arc while projection axis~3 is perpendicular to it, and projection axis~2 is between 1 and 3.
Fig.~\ref{fig:projection_rationale} (b) presents the projection results on these three axes.
We can observe that, different projections have different capabilities in terms of sensing respiration: while projection 3 has very little fluctuation, projection 1 and 2 have clear periodical fluctuation patterns corresponding to inhalation and exhalation.

We can see that compared to traditional approach~\cite{droitcour2004range} which simply selects a better one from I and Q (weight is either 0 or 1), we tune the weight for combination in a fine-grained manner by changing the projection parameter $\theta$ from 0 to $2\pi$ at a chosen step size $\frac{\pi}{n}$. Our approach will generate $2n$ combination candidates, which is much larger than that of the traditional approach.
The selection between I and Q can be viewed as a combination of I/Q with $\theta=0$ or $\theta=\frac{\pi}{2}$. Thus the traditional approach is actually a small subset of our approach.
The much larger number of combination candidates make our approach outperform the previous work.

\subsubsection{Selection from Multiple Candidates}
\label{sec:pattern_extraction:2:2}

After generating multiple combination candidates, we need to select the best one from these candidates to generate the respiration pattern.
Our insight comes from that the target's respiration rate can be assumed a constant during a short period of time \cite{zeng2018fullbreathe, yue2018extracting}.
Therefore, the periodicity of the signal pattern during a short period can represent its capacity to sense respiration, which can be used to select the best one from the candidates.

Here, we adopt the short term breathing-to-noise ratio (BNR) proposed in~\cite{yue2018extracting} to measure the periodicity of a combination candidate, which is defined as the ratio of respiration energy to the overall energy.
We compute BNR by first taking FFT of the combination candidate. In our computation, the window length of projection is set to 12 seconds which corresponds to 1200 samples.
Next, we find the FFT bin with maximal energy within the human respiration range (10\,bpm to 37\,bpm).
Here, we increase the number of samples to 8192 by means of zero-padding which appends the time-domain signal with 6992 zero-value samples~\cite{stoica2005spectral}. 
The zero-padding of course can not improve the spectral resolution, however, it can reveal finer details in the spectrum so that the FFT bin with maximal energy can be more easily located~\cite{kay1988modern}.
Then BNR is calculated by dividing that bin's energy by the energy sum of all FFT bins.
At last, we select the one that has the maximal BNR value among all the combination candidates.

Our selection strategy based on periodicity performs better than the commonly used strategy that is based on variance\cite{wang2016human, wang2017phasebeat}.
When the target is near to the transceiver pair, both strategies can effectively extract the respiration pattern.
However, when the target is further away from the transceivers, even the CSI ratio becomes noisy and it is no longer a circular arc any more. Then the selection based on variance may fail since the respiration-caused signal variation is now smaller than the noise level and in this case the variance is mainly caused by environmental noise
~\cite{wang2015understanding}.
Differently, our selection strategy based on periodicity still tries its best to combine I/Q for extracting the quasi-periodical respiration pattern during a short period.

\begin{figure*}[t]
	\begin{minipage}[t]{1\linewidth}
		\centering
		\includegraphics[width=1\textwidth]{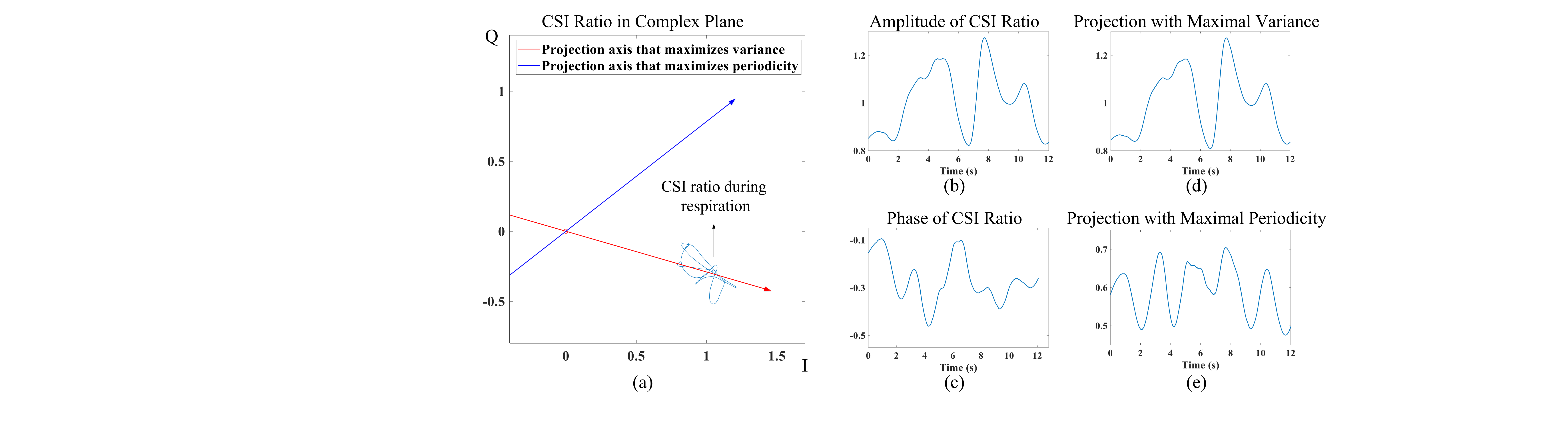}
		\caption{Comparison of different respiration pattern extraction approaches when a subject is far away from the transceivers: (a) CSI ratio in complex plane is no longer a circular arc; (b) amplitude; (c) phase; (d) projection with maximal variance; (e) projection with maximal periodicity.
			We can see clear and periodical respiration pattern in (e).
		}
		\label{fig:selection_cmp}
	\end{minipage}	
\end{figure*}

Fig.~\ref{fig:selection_cmp} compares different respiration pattern extraction methods when a target breathes naturally far away from the transceivers and the ground-truth respiration rate is 25\,bpm.
Fig.~\ref{fig:selection_cmp} (b) and (c), which employ the amplitude and phase of the CSI ratio, have no clear rhythmical pattern caused by respiration.
In Fig.~\ref{fig:selection_cmp} (d), the projection has the maximal variance.
However, the rhythmical pattern is still not clear and respiration can hardly be detected.
In contrast, the projection with maximal periodicity in Fig.~\ref{fig:selection_cmp} (e) shows clear respiration pattern, which matches the ground truth well.
Obviously, our approach outperforms the selection of a better one from amplitude/phase \cite{zeng2018fullbreathe} and the I/Q combination with maximal variance, thus significantly increasing the respiration sensing accuracy and range.
\section{The FarSense System}
\label{sec:system}

\begin{figure*}[t]
	\begin{minipage}[t]{0.96\linewidth}
		\centering
		\includegraphics[width=0.96\textwidth]{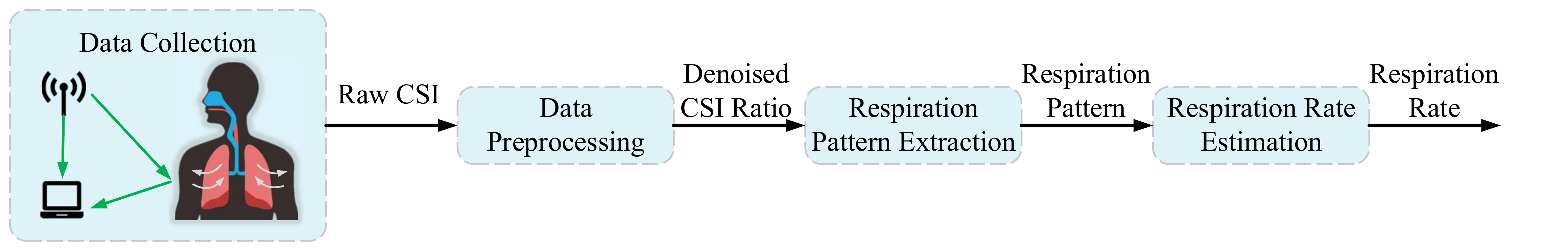}
		\caption{System architecture of FarSense.}
		\label{fig:system_overview}
	\end{minipage}		
\end{figure*}

In this section, we present our design and implementation of a real-time respiration monitoring system named FarSense.
The FarSense system consists of four basic modules: Data Collection, Data Preprocessing, Respiration Pattern Extraction and Respiration Rate Estimation, as shown in Fig.~\ref{fig:system_overview}.

\subsection{Data Collection}
\label{sec:system:1}
In this module, we collect CSI data from two antennas at the receiver using the CSI tool \cite{halperin2011tool} developed by Halperin which collects the CSI samples for each received packet.
Here, we configure the WiFi card to run at a central frequency of 5.24\,GHz with a bandwidth of 20\,MHz.
Note that the Intel 5300 WiFi card provides CSI on 30 sub-carriers out of a total of 56 sub-carriers for 20\,MHz bandwidth.
The sampling rate of CSI is set to 100\,Hz in our system.
Although a sampling rate of 10\,Hz is enough to capture the change of CSI caused by respiration~(smaller than 1\,Hz), we set a higher sampling rate for higher respiration rate estimation resolution using autocorrelation in Sec.~\ref{sec:system:4}.

\subsection{Data Preprocessing}
\label{sec:system:2}
In this module, for each sub-carrier, we first divide the two complex CSI readings from the two antennas at the same receiver to obtain the CSI ratio which is still a complex value.

Next, we introduce the motion detector proposed in \cite{li2018training} to flag periods of time when there are large motions and exclude these time periods for sensing.
This is because when the human target is at high mobility, the CSI change due to minute chest movement is overwhelmed by the large motions so that FarSense system will not be able to accurately monitor respiration.
Different from \cite{li2018training} that feeds the conjugate multiplication of CSI from two antennas into the speed spectrum estimator, we feed the CSI ratio into it to robustly determine whether the human target is stationary or non-stationary (i.e., whether there are large motions).

At last, we apply the Savitzky-Golay filter to smooth the CSI ratio data of each sub-carrier during these stable time periods and use them for further processing.

\subsection{Respiration Pattern Extraction}
\label{sec:system:3}
In this module, we extract respiration patterns of 30 sub-carriers.
For each sub-carrier $i$, we go through all the projection axis parameter $\theta$ (i.e., $0, \frac{\pi}{50}, \frac{2\pi}{50}, ..., \frac{99\pi}{50}$ at a step size of $\frac{\pi}{50}$) to generate 100 candidates and select one that maximizes BNR as the final extracted respiration pattern, as shown in Sec.~\ref{sec:pattern_extraction:2}.
For the laptop (DELL Precision 5520 with Intel Xeon E3-1505M v6 and 8\,GB RAM) we use for data processing, the time cost of all 30 sub-carriers for searching $\theta$ from 0 to $2\pi$ with a given step $\frac{\pi}{50}$ is about 0.54\,s, which is acceptable for a real-time respiration monitoring system.
We can further reduce the time cost by increasing the step size of searching.

\subsection{Respiration Rate Estimation}
\label{sec:system:4}
In this module, we estimate respiration rate by combining results from multiple sub-carriers.
We adopt the autocorrelation method which has shown to perform well in low SNR scenario~\cite{droitcour2006non}. 
The periodicity of respiration presents us this unique opportunity to obtain peaks with autocorrelation. We first apply autocorrelation on the respiration pattern of each sub-carrier.  Then we combine the autocorrelation results from multiple sub-carriers to obtain a final respiration rate estimation.

\subsubsection{Applying Autocorrelation on Respiration Pattern}
\label{sec:system:4:1}
First, for each sub-carrier, we calculate the autocorrelation of its respiration pattern. 
The autocorrelation function describes the similarity of a signal to a shifted version of itself.
According to \cite{box2015time}, for sub-carrier $i$, the autocorrelation $\mathbf{r}_i(k)$ of a time series of respiration pattern $\mathbf{y}_i$ for a shift of $k$ samples is defined as:
\begin{equation}
\label{equation:acf}
	\mathbf{r}_i(k) = \dfrac{\sum\limits_{t=k+1}^T (\mathbf{y}_i(t)-\overline{\mathbf{y}_i})(\mathbf{y}_i(t-k)-\overline{\mathbf{y}_i})}
	{\sum\limits_{t=1}^T (\mathbf{y}_i(t)-\overline{\mathbf{y}_i})^2}
\end{equation} 
where $T$ is the total length of $\mathbf{y}_i$, $\overline{\mathbf{y}_i}$ is the mean of $\mathbf{y}_i$ and $k=0,...,T-1$ is number of samples shifted. 

\begin{figure*}[t]
	\begin{minipage}[t]{1\linewidth}
		\centering
		\includegraphics[width=1\textwidth]{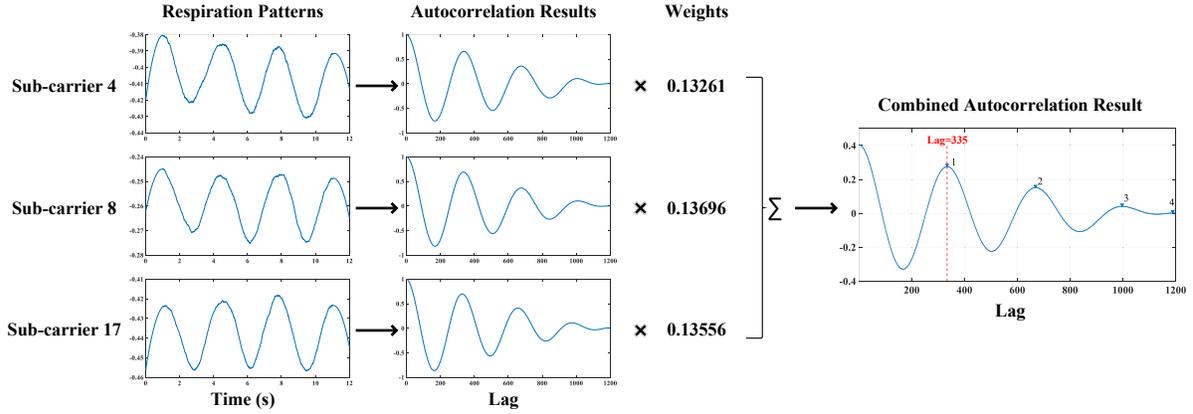}
		\caption{
			Example of multiple sub-carriers combining.
			In this case, we use the weighted sum of three selected sub-carriers' autocorrelation results to obtain a final one.
			The estimated respiration rate is $\frac{60}{335/100}=17.9$\,bpm.
		}
		\label{fig:msc}
	\end{minipage}	
\end{figure*}

\subsubsection{Multiple Sub-carriers Combining}
\label{sec:system:4:2}
We now combine the sub-carriers by employing a weighted sum of each sub-carrier's autocorrelation result. The weight of each sub-carrier is the respiration pattern's BNR value.
With extensive experiments, we find that, when the subject is far away from the WiFi transceivers, there are occasionally some so-called “bad” sub-carriers whose respiration patterns are chaotic even after we apply our proposed methods. We believe this is because each sub-carrier experiences different multipath fading and shadowing effects, and the CSI ratios of different sub-carrier have different sensitivity to subtle respiration motions.
Thus, we do not include all the sub-carriers but only those "good" ones. 
Those "bad" sub-carriers which have BNR values smaller than a pre-defined threshold will be excluded from combination.
Let the maximal BNR among all 30 sub-carriers be $\varepsilon$, we just include those sub-carriers whose BNR is larger than $0.7\varepsilon$ for combination.\footnotemark[3]
\footnotetext[3]{We tested different thresholds with large amounts of data and empirically chose 0.7 for sub-carrier selection.}
And the final combined autocorrelation result can be represented as:
\begin{equation}
\label{equation:msc}
\mathbf{r}_{msc} = \sum\limits_{i \in S} BNR_i \times \mathbf{r}_i
\end{equation}
where $S$ is the set of sub-carriers whose BNR is larger than $0.7\varepsilon$, $BNR_i$ is the BNR of sub-carrier $i$ and $\mathbf{r}_i$ is the autocorrelation result of sub-carrier $i$.
The first peak of $\mathbf{r}_{msc}(k) (k=0,...,T-1$) is the component describing the periodicity of respiration \cite{jeyhani2017comparison}. And the shift of the first peak divided by the sampling rate is the estimated period for one respiration cycle.

Fig.~\ref{fig:msc} shows an example of respiration rate estimation by combining multiple sub-carriers.
Three sub-carriers (i.e., 4, 8 and 17) are selected to participate in the weighted sum operation.
As shown in the final combined autocorrelation result, the lag~(shift) of the first peak labeled with 1 is 335.
For a sampling rate of 100\,Hz, the estimated respiration rate is calculated as $\frac{60}{335/100}=17.9$\,bpm.
\section{Evaluation}
\label{sec:evaluation}

In this section, we conduct comprehensive experiments to evaluate the performance of FarSense with commodity WiFi devices.
In Sec.~\ref{sec:eva:1}, we describe the experimental setup.
In Sec.~\ref{sec:eva:2}, we compare FarSense with the state-of-the-art approaches in terms of sensing range. 
In Sec.~\ref{sec:eva:3}, we evaluate the robustness of FarSense in challenging real-life scenario such as when the WiFi AP is located in another room far away from the target.

\begin{figure*}[t]
	\begin{minipage}[t]{0.41\linewidth}
		\centering
		\includegraphics[width=1\textwidth]{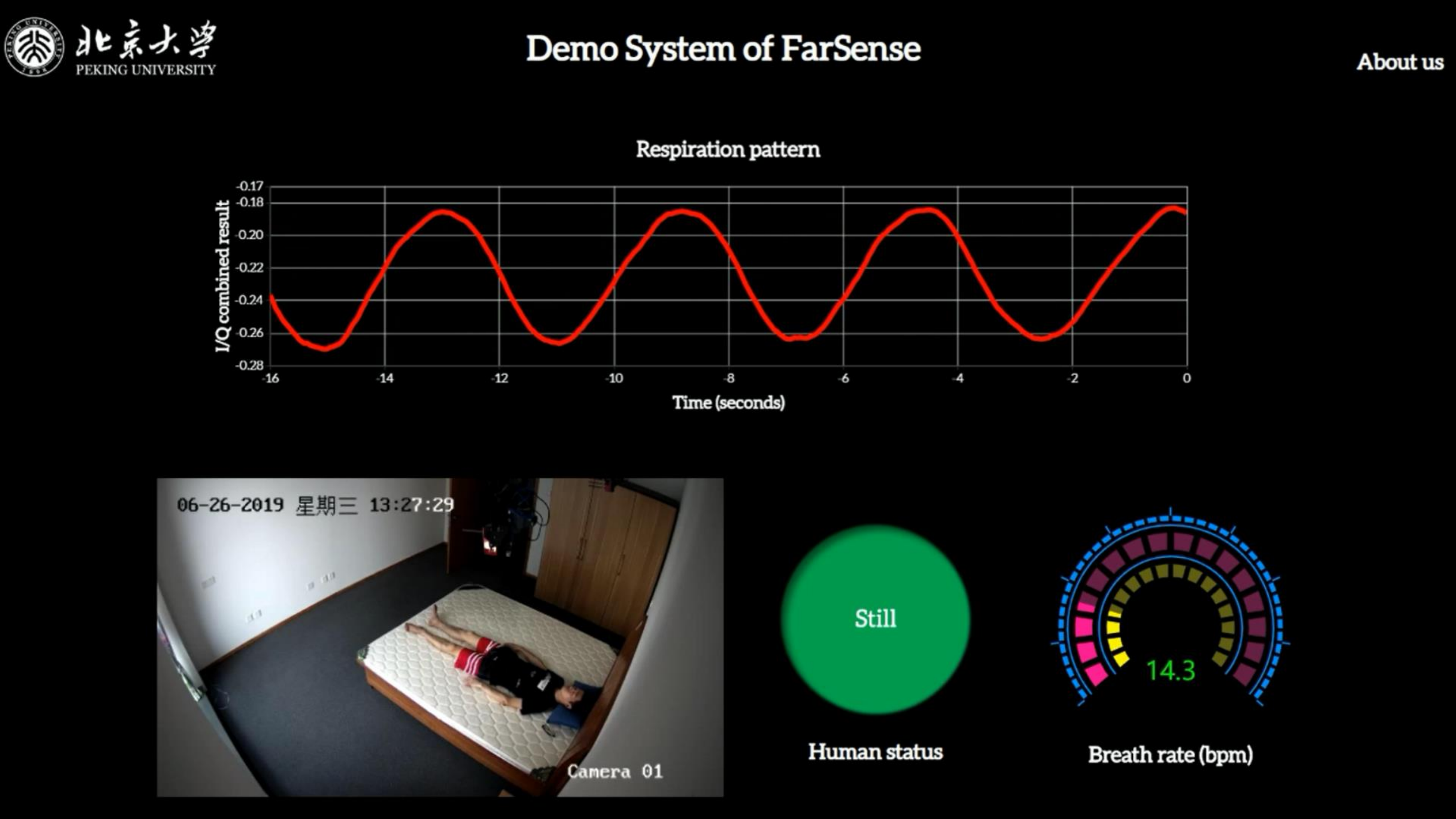}
		\caption{Graphical user interface of FarSense.
		}
		\label{fig:web}
	\end{minipage}	
	\hspace{2pt}
	\begin{minipage}[t]{0.57\linewidth}
		\centering
		\includegraphics[width=1\textwidth]{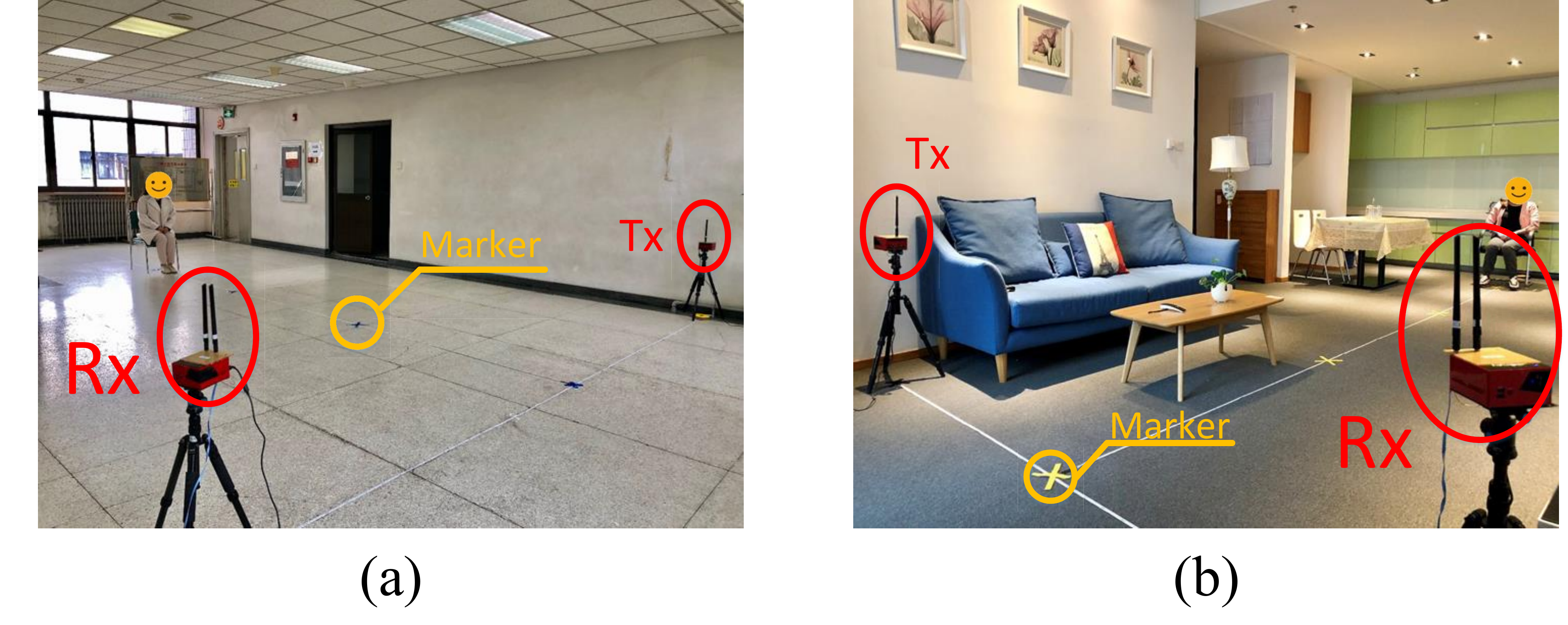}
		\caption{Comparison with the state-of-the-art approaches in two different experimental environments: (a) a large corridor; (b) a typical home.
		}
		\label{fig:cmp_setting}
	\end{minipage}
\end{figure*}

\subsection{Experimental Setup}
\label{sec:eva:1}
As shown in Sec.~\ref{sec:model:3}, we employ a pair of GIGABYTE mini-PCs equipped with cheap Intel 5300 WiFi cards as transceivers where one antenna is equipped at the transmitter~(Tx) and two antennas are equipped at the receiver~(Rx).
The carrier frequency of the WiFi channel is set as 5.24\,GHz and the transmitter broadcasts 100 packets one second.
We collect CSI data at the Rx using the CSI tool~\cite{halperin2011tool} and process it with MATLAB at a DELL Precision 5520 laptop (Intel Xeon E3-1505M v6, 8\,GB RAM) in real time.
Fig.~\ref{fig:web} shows the graphical user interface~(GUI) of FarSense, which consists of four components: (1) the respiration pattern of the sub-carrier that has the largest BNR; (2) the real-time video of the test environment captured by a camera; (3) the human status (stationary or non-stationary); and (4) the estimated respiration rate.
Note that if the human status is non-stationary, the GUI will not display the value of respiration rate.
Different from the lab-controlled ground-truth collection approaches which ask subjects to breathe to a metronome~\cite{kaltiokallio2014non, ravichandran2015wibreathe}, we collect the ground-truth respiration rates when a subject breathes naturally with a commercial device (Neulog Respiration Monitor Belt
logger sensor NUL-236 \cite{kam2017compact}).
In the experiments, we recruit 12 people and collect a total of 197 hours of CSI data in different environments where each subject sits on a chair or lies in a bed. 

\subsection{Comparison with Previous Approaches}
\label{sec:eva:2}

In this subsection, we compare FarSense with two state-of-the-art WiFi-based respiration sensing approaches \cite{wang2016human, zeng2018fullbreathe} in terms of sensing range.
To quantitatively understand the effective sensing range, we adopt the metric of detection rate when a subject breathes at locations with different distances to the transceivers.
As shown in previous section, the distance from a subject to the transceivers is defined as the average distance from the subject to the transmitter and receiver.
The detection rate at a certain distance is defined as $\frac{N_{detected}}{N_{all}}$, where $N_{detected}$ is the number of CSI measurements whose respiration rate estimation is very close to the ground truth (i.e., the absolute error is less than 0.5\,bpm) and $N_{all}$ is the total number of collected CSI measurements. 
Then, we define the sensing range of a respiration sensing system as the maximal distance at where the detection rate is higher than 95\%.
Obviously, a higher detection rate means a larger sensing range.

\subsubsection{Baseline Approaches}
\label{sec:eva:2:1}
We first introduce these two state-of-the-art approaches as follows:
\begin{itemize}
	\item \textbf{HRD.}\footnotemark[4]
	\footnotetext[4]{For convenience, we use the term "HRD" (Human Respiration Detection) to represent the respiration detection system proposed in \cite{wang2016human}.}
	HRD \cite{wang2016human} uses the CSI amplitude for respiration sensing.
	HRD first applies the Hampel filter and a moving average filter to remove outliers and high-frequency noise. Then HRD employs a threshold-based method to select sub-carriers that have larger variance of CSI amplitude and estimates the respiration rate with peak detection method.
	
	\item \textbf{FullBreathe.} FullBreathe \cite{zeng2018fullbreathe} uses the conjugate multiplication~(CM) of CSI readings from two adjacent antennas at the same sub-carrier for sensing. FullBreathe first applies the Savitzky-Golay filter to smooth the noisy raw CM. FullBreathe then selects either the amplitude or phase which achieves better performance for sensing. FullBreathe also estimates the respiration rate by peak detection.
\end{itemize}

\subsubsection{Experimental Settings}
\label{sec:eva:2:2} 
We conduct experiments in two different environments, as shown in Fig.~\ref{fig:cmp_setting} (a) and (b): one is a large corridor and the other is a typical home environment with furniture and electrical appliances which has rich multipath. 
In each environment, we deploy a pair of transceivers and ask a subject to sit in the chair located on the perpendicular bisector of the LoS path of the transceivers.
We place markers on the floor to label the distances (i.e., 0\,m, 2\,m, 4\,m and 6\,m) to the LoS path and record CSI data when a subject breathes naturally.
We vary the distance between the subject and the LoS path from 2.5\,m to 5\,m at a step size of 0.1\,m.
Since the subject is on the perpendicular bisector of the LoS path, following simple geometry, we can obtain the distance from the subject to a transceiver pair
$\sqrt{({\frac{LoS}{2}})^{2}+l^2}$, where $l$ is the distance between the subject and the LoS path.
We further vary the distance between the transmitter and receiver from 3\,m to 5.5\,m at a step size of 0.1\,m so we record CSI data at different distances to the transceivers ranging from $\sqrt{({\frac{3}{2}})^{2}+2.5^2}=2.9$\,m to $\sqrt{({\frac{5.5}{2}})^{2}+5^2}=5.7$\,m.
To see the effect of environment changes, we also move the furniture randomly to change the multipath.

\subsubsection{Experimental Results}
\label{sec:eva:sensing_range}
To clearly visualize the comparison results, we show them from two aspects: one is the detailed respiration patterns in one of the above settings and the other is the overall detection rate of each approach with varying distances between the target and the transceivers.

\begin{figure*}[t]
	\begin{minipage}[t]{0.63\linewidth}
		\centering
		\includegraphics[width=0.99\textwidth]{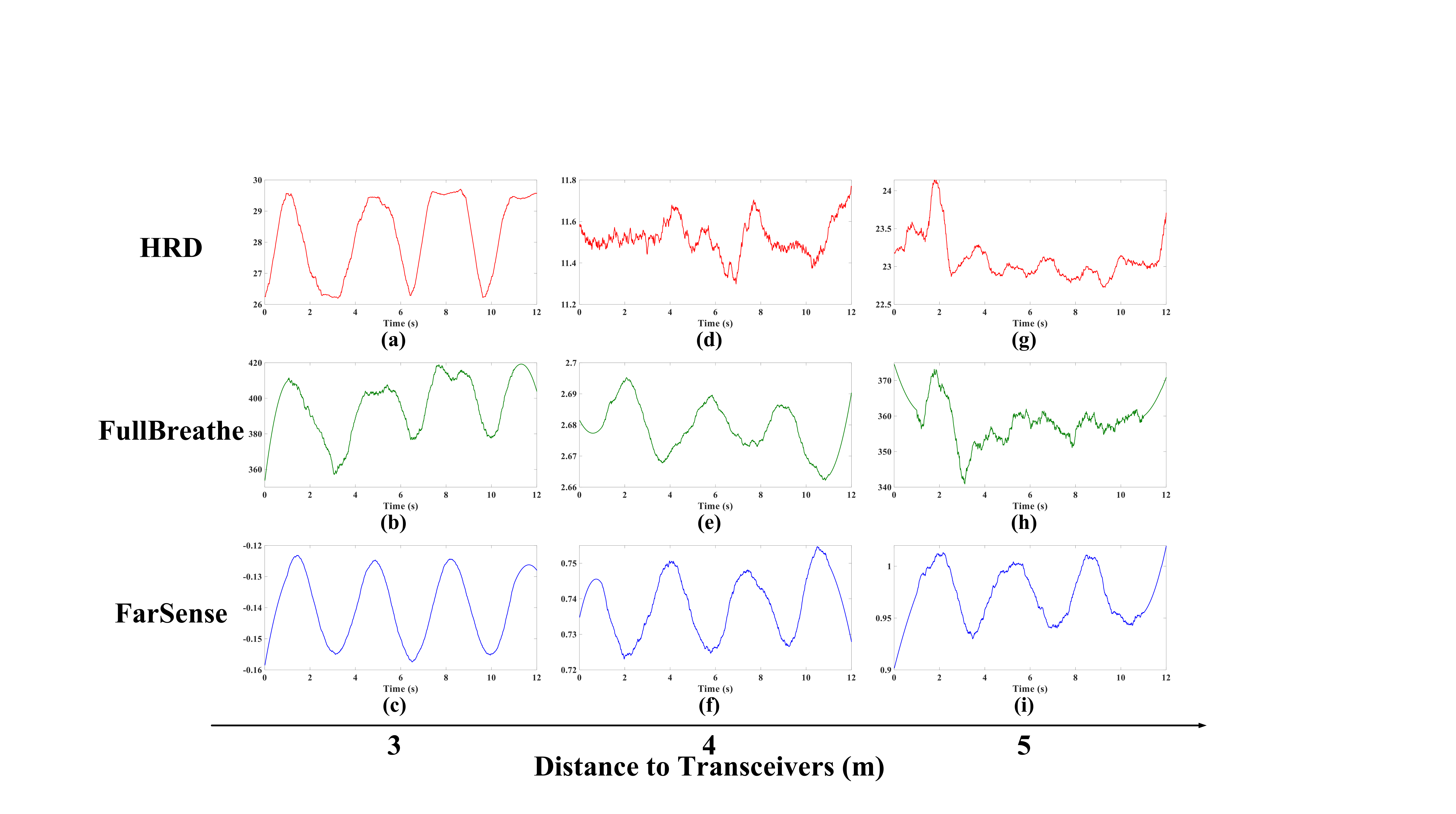}
		\caption{The detailed respiration patterns of each approach at three different distances to the transceivers.
		At the distance of 5\,m, FarSense can still effectively sense respiration while the other two approaches fail.
		}
		\label{fig:cmp_pattern}
	\end{minipage}
	\hspace{2pt}
	\begin{minipage}[t]{0.34\linewidth}
		\centering
		\includegraphics[width=0.99\textwidth]{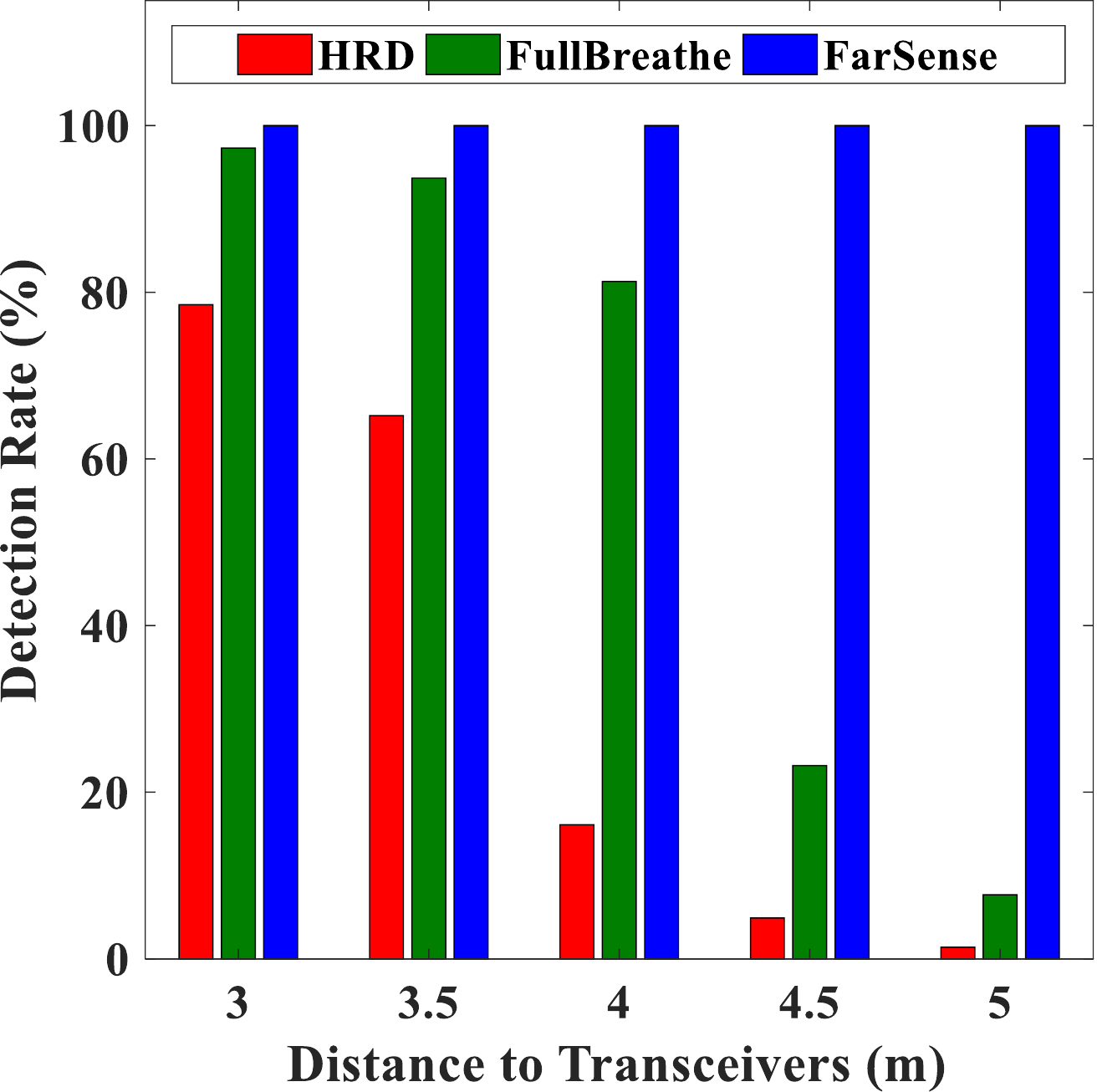}
		\caption{The overall detection rate of each approach versus distance to the transceivers.
		}
		\label{fig:cmp_result}
	\end{minipage}		
\end{figure*}

\textbf{Respiration Pattern.}
Fig.~\ref{fig:cmp_pattern} presents the detailed respiration patterns of the three approaches at different distances to the transceivers (i.e., 3\,m, 4\,m and 5\,m) in a typical home with the transmitter and receiver separated by 3\,m.
The ground truth of respiration rate is 18.2\,bpm, which means $\frac{18.2}{60/12} \approx 3.5$ peaks/valleys in a 12-second window.
When the subject is 3\,m away from the transceivers, we can observe the clear respiration patterns with all three approaches. We can see that among the three clear patterns, FarSense still achieves the clearest respiration pattern.
When the distance is increased to 4\,m, as shown in Fig.~\ref{fig:cmp_pattern} (d), the pattern obtained with HRD is full of noise, and we can hardly extract respiration rate from it.
However, FullBreathe (Fig.~\ref{fig:cmp_pattern} (e)) and FarSense (Fig.~\ref{fig:cmp_pattern} (f)) still have clear fluctuations, which match the ground truths.
If the subject moves 5\,m away from the transceivers, as shown in Fig.~\ref{fig:cmp_pattern} (g)-(i), both HRD and FullBreathe fail to extract clear respiration patterns while FarSense still performs well.
Under other settings specified in Sec.~\ref{sec:eva:2:2}, FarSense always outperforms the other two state-of-the-art systems and we skip the similar results here.

\textbf{Overall Detection Rate.}
Fig.~\ref{fig:cmp_result} presents the overall detection rates of the three approaches with different distances away from the transceivers ranging from 3\,m to 5\,m at a step size of 0.5\,m.
The figure shows that our detection rate remains 100\% even when the subject is 5\,m away from transceivers, while the detection rate of HRD and FullBreathe drop to 1.4\% and 7.7\%, respectively.
Obviously, at a greater distance, both HRD and FullBreathe can hardly detect the human respiration, achieving very low detection rates.
According to the definition of sensing range in Sec.~\ref{sec:eva:2}, we get the sensing range for HRD (less than 2.9\,m), FullBreathe (3.7\,m) and FarSense (larger than 5.7\,m).
The experimental results demonstrate the effectiveness of FarSense in terms of increasing the sensing range without sacrificing the accuracy.

\subsection{Performance in Challenging Real-life Scenarios}
\label{sec:eva:3}

In this subsection, we evaluate FarSense's performance in challenging real-life scenarios.
In Sec.~\ref{sec:eva:3:1}, we conduct experiments when the subject is located 6\,m to 9\,m away from the transceivers to evaluate the range limit of our system.
Apart from the LoS scenarios, in Sec.~\ref{sec:eva:3:2}, we evaluate the FarSense's performance when the subject has non-LoS path with the transmitter (or receiver)--the WiFi transmitter is placed in a different room from the receiver with a wall in between and the subject is located either in the transmitter side or in the receiver side.
In addition to sitting scenario, in Sec.~\ref{sec:eva:3:3}, we evaluate FarSense when the subject lies in bed with different sleeping postures while both transceivers are mounted far away on the ceiling.

\subsubsection{Sitting Far from the Transceivers}
\label{sec:eva:3:1} 

\begin{figure*}[t]
	\begin{minipage}[t]{0.44\linewidth}
		\centering
		\includegraphics[width=0.99\textwidth]{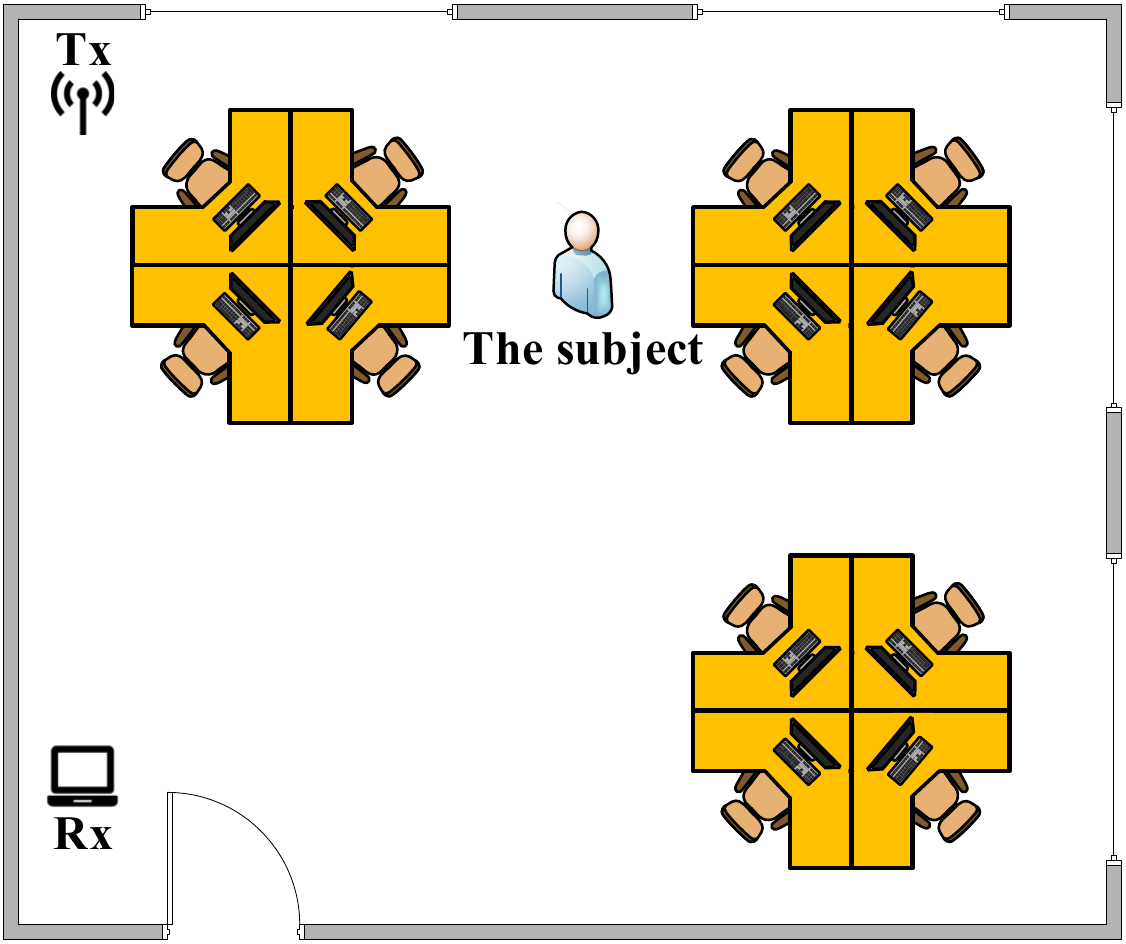}
		\caption{Experimental settings in a large office room (7.5\,m $\times$ 9\,m).
		}
		\label{fig:far_setting}
	\end{minipage}
	\hspace{4pt}
	\begin{minipage}[t]{0.37\linewidth}
		\centering
		\includegraphics[width=0.99\textwidth]{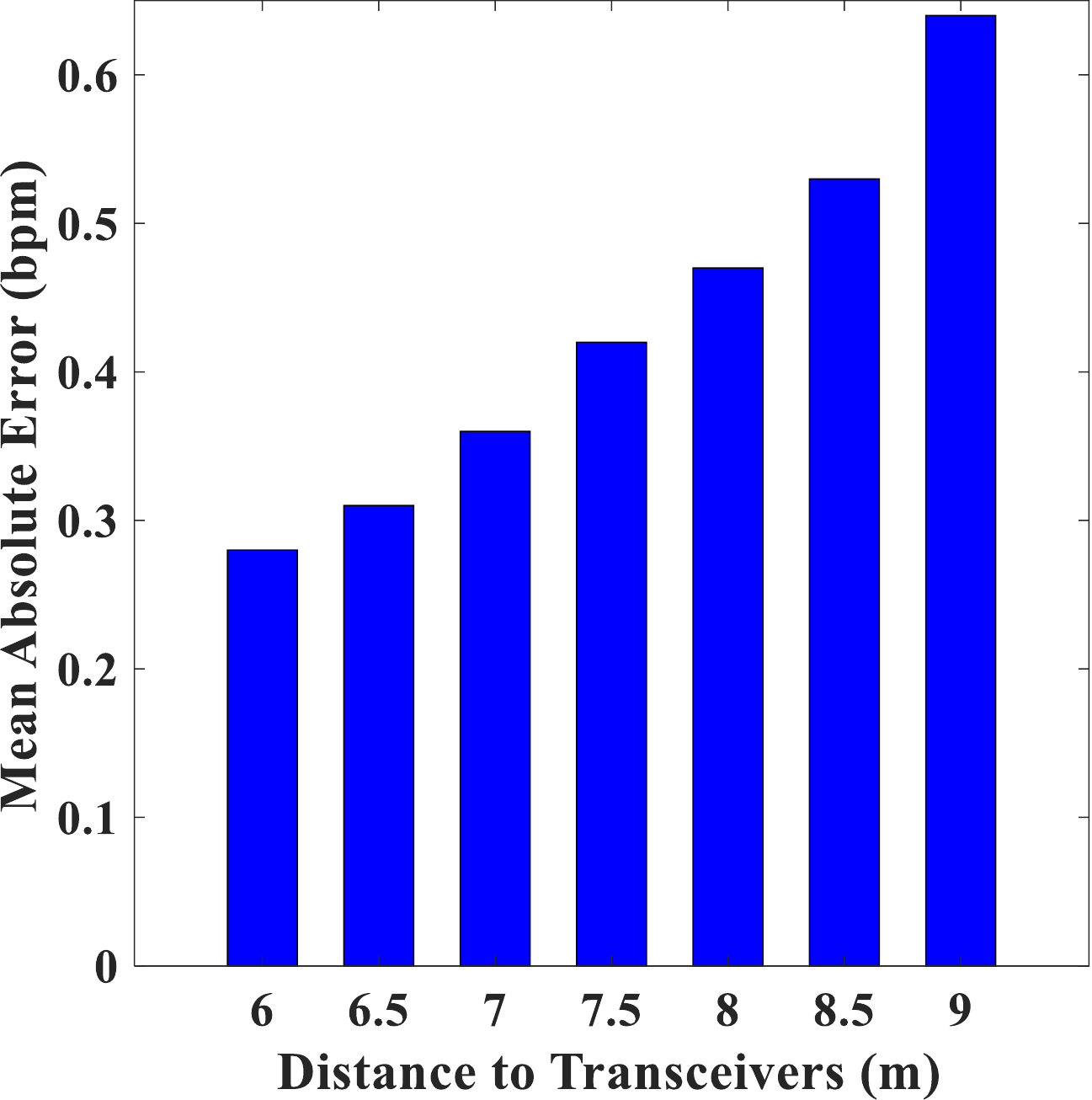}
		\caption{The mean absolute error of respiration rate versus distance to the transceivers.
		}
		\label{fig:far_result}
	\end{minipage}		
\end{figure*}

We now evaluate FarSense's performance when the subject sits beyond 5\,m to the transceivers.
As shown in Fig.~\ref{fig:far_setting}, we conduct the experiments in a large office room (7.5\,m $\times$ 9\,m), where rich multipath exists due to a large number of furniture and electronic appliances.
In the experiments, the transceivers are placed in the corner with a LoS path length of 6.8\,m, and the subject sits on a barstool with a height of 0.95\,m, breathing naturally.
We randomly move the subject's position so that he/she is not just located on the perpendicular bisector of the LoS path.
We measure the distances from the subject to the transmitter and receiver with a laser meter \cite{pu2004laser} and average these two distances to obtain the distance from the subject to the transceivers.

We plot the mean absolute error of respiration rate as a function of distance from 6\,m to 9\,m in Fig.~\ref{fig:far_result}.
The figure shows that the mean absolute error of FarSense is 0.28\,bpm at 6 meters and slightly increases to 0.64\,bpm even when the subject is 9 meters away from the transceivers.
When the distance from the subject to transceivers increases, the mean absolute error increases, too.
This is because the signal reflected off human target is further attenuated when the distance is increased.
We also observe that, the mean absolute error is still less than 0.5\,bpm when the distance is 8\,m.
That is to say, FarSense can reliably sense human respiration at 100\% detection rate (less than 0.5\,bpm) even when the subject is 8\,m away from the transceivers.
According to the definition of sensing range in Sec.~\ref{sec:eva:2}, the sensing range of FarSense is larger than 8\,m. Compared to HRD (less than 2.9\,m) and FullBreathe (3.7\,m), the sensing range of FarSense is increased by more than $\frac{8-3.7}{3.7}=116$\%.

\subsubsection{NLoS Scenarios}
\label{sec:eva:3:2}

\begin{figure*}[t]
	\begin{minipage}[t]{0.87\linewidth}
		\centering
		\includegraphics[width=0.99\textwidth]{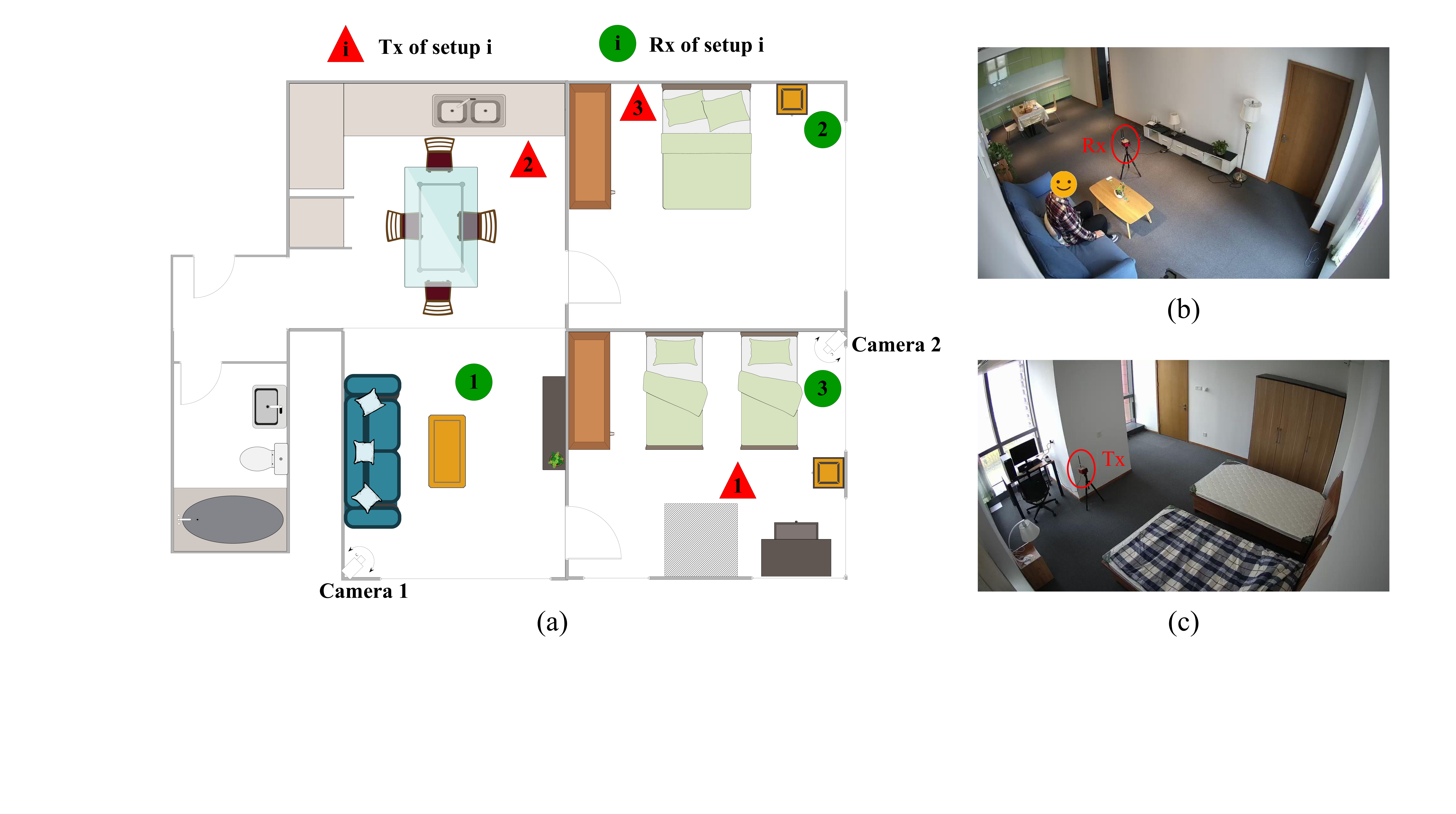}
		\caption{The Experiments for NLoS scenarios are conducted in a typical home: (a) the floor plan of the home with three different transceiver setups, namely setup 1, 2 and 3; (b) the photo for setup 1, which is captured by camera 1; (c) the photo for setup 1, which is captured by camera 2.
		}
		\label{fig:nlos_setting}
	\end{minipage}		
\end{figure*}

To evaluate the ability of FarSense to monitor human respiration in NLoS scenarios, we conduct experiments when the WiFi transmitter is placed in a different room from the receiver with a wall in between, and the subject is located either in the transmitter side or in the receiver side.
As shown in Fig.~\ref{fig:nlos_setting} (a), we conduct the experiments in a typical home with three different transceiver setups, namely setup~1, 2 and 3.
For each setup, the subject sits on the sofa, in the bed or a chair, and is about 5\,m to the transceivers.
Fig.~\ref{fig:nlos_setting} (b) and (c) present the two photos of deployment for setup 1, which is captured by one camera in the living room (camera 1) and another camera in the bedroom (camera 2), respectively.
In this setup, the subject sitting on the sofa is in the same room with the receiver while the transmitter is in the bedroom with a 10cm-thick wall between the two rooms.

For all the experiments in NLoS scenarios, the achieved mean absolute error is as small as 0.34\,bpm.
The results indicate that FarSense can reliably monitor human respiration even in the challenging NLoS scenario with a wall in between.
We also observe that the mean absolute error at the same distance of 5\,m for the NLoS scenarios is slightly higher than that for the LoS scenarios (0.23\,bpm) in Sec.~\ref{sec:eva:2}.
This is because the signal reflected off the human target becomes even weaker after penetrating the wall. We also want to point out that the other two state-of-the-art systems fail to work in these NLoS scenarios.

\subsubsection{Sleeping in Different Postures}
\label{sec:eva:3:3}
\begin{figure*}[t]
	\begin{minipage}[t]{0.45\linewidth}
		\centering
		\includegraphics[width=0.99\textwidth]{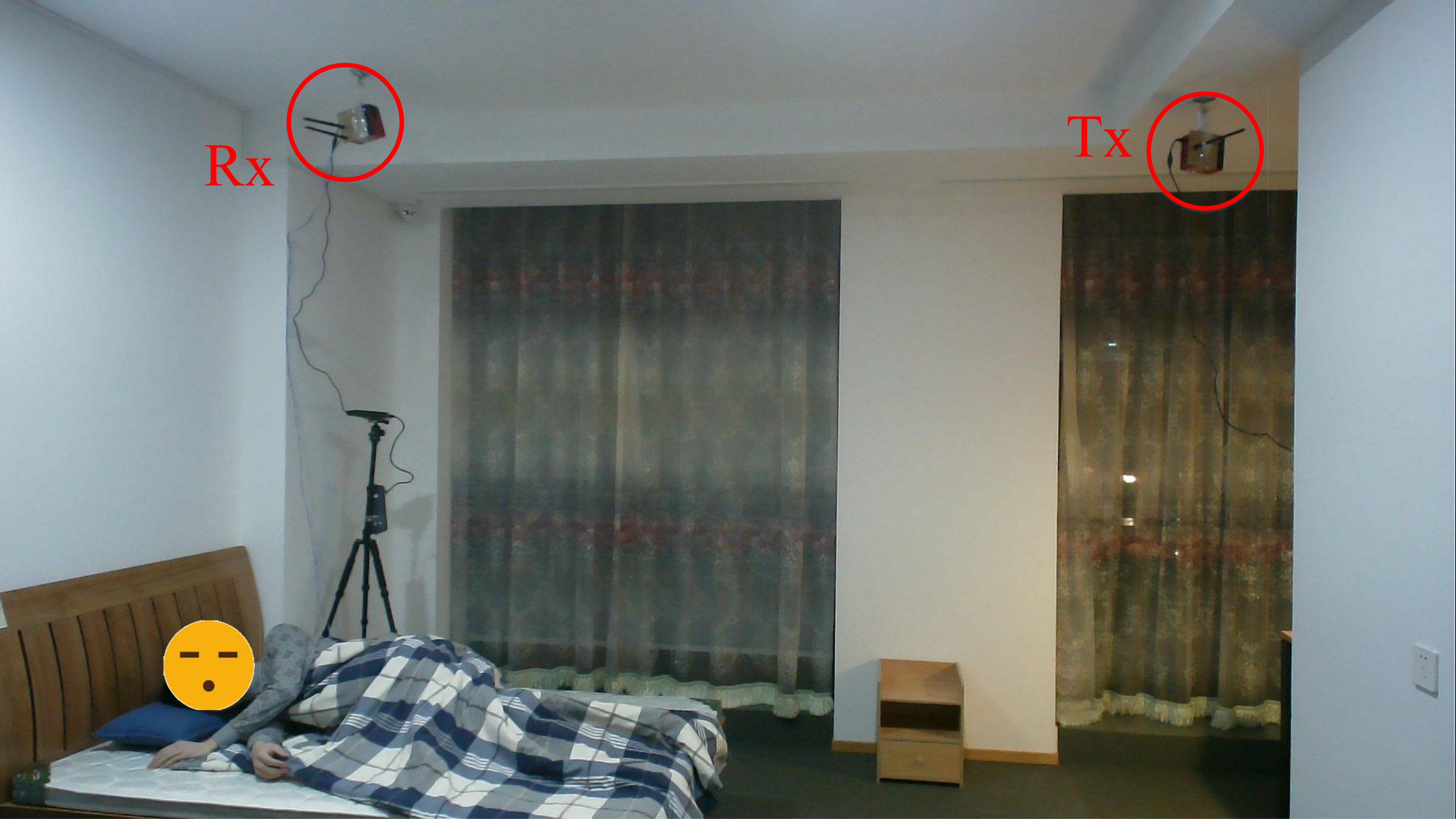}
		\caption{Experimental settings for sleeping scenarios.
		}
		\label{fig:sleep_setting}
	\end{minipage}
	\hspace{4pt}
	\begin{minipage}[t]{0.35\linewidth}
		\centering
		\includegraphics[width=0.99\textwidth]{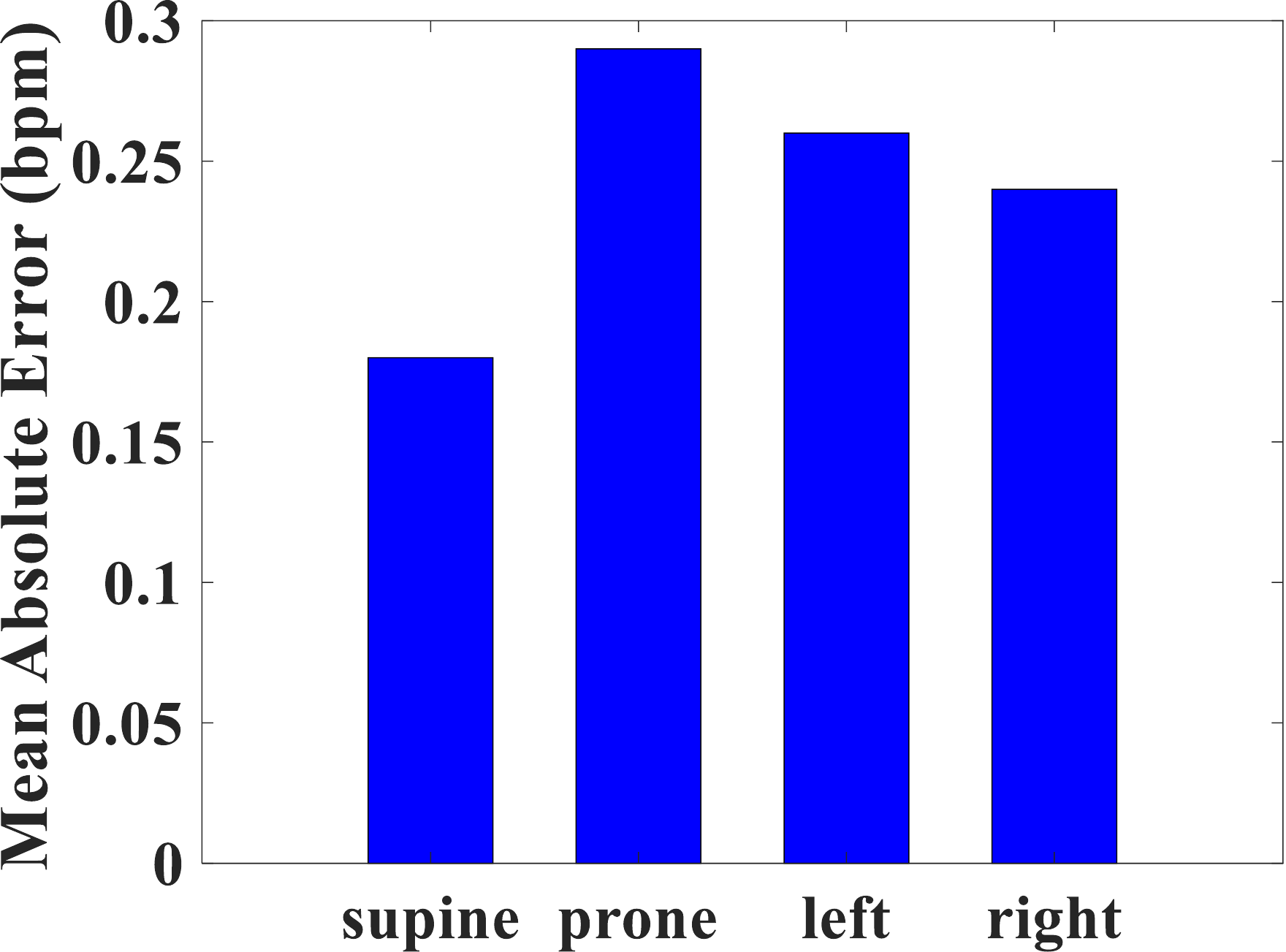}
		\caption{Mean absolute error of respiration rate versus the sleeping postures.
		}
		\label{fig:sleep_result}
	\end{minipage}		
\end{figure*}

We conduct experiments when the subject sleeps in the bed with four different postures, i.e., supine, prone, left and right.
Different from the lab-controlled settings in \cite{liu2015tracking, wang2016human} where the transceivers are placed close at two sides of the bed, we mount the transceivers on the ceiling which is a deployable setup in real indoor environments, far away from the subject. To make it even more challenging, the subject is covered with a quilt.
As shown in Fig.~\ref{fig:sleep_setting}, the antennas of the transceivers are placed horizontally.
In this setting, the LoS path length is 3.9\,m and the distance from the subject to the transceivers mounted on the ceiling is 3.8\,m.

Fig.~\ref{fig:sleep_result} shows the mean absolute error of respiration rate for different sleeping postures.
As shown in the figure, no matter at what posture, the mean absolute error is always less than 0.3\,bpm, which demonstrates the effectiveness and robustness of FarSense.
As can be seen, the respiration rate is more accurate in supine posture than in other postures.
We believe this is because when the subject's chest faces the transceivers in supine posture, the chest displacement induces larger signal fluctuations and and thus achieves more accurate estimation.
\section{Discussions}
\label{sec:discussions}

In this work, we focus on designing a single-person respiration detection system with commodity WiFi devices, which achieves a much further  sensing range than the state-of-the-art WiFi-based systems. We briefly discuss the limitation and future direction of our work below.

\subsection{Through the Wall Twice}
\label{sec:discussions:1}
FarSense works well when one of the transceivers is in another room. However, when both WiFi transceivers are located in another room, FarSense has difficulty to sense human respiration because now the weak reflected signal from the human chest will need to go through the wall twice~\cite{adib2013see}.
A potential solution is to null out the strong static LoS signal and other reflections as shown in \cite{adib2013see}  for FMCW radar.

\subsection{Multiple Subjects}
\label{sec:discussions:2}
In real life, there are scenarios when multiple subjects are in the same room, e.g., a couple is sharing a same bed.
It is a well-known challenge to separate the signals reflected off multiple targets and achieve multi-target sensing with cheap commodity hardware.
Basically there are two ways to address the challenging multi-target sensing issue with WiFi. One is to increase the bandwidth and the other is to increase the number of antennas at the WiFi transceivers. 

For example, with a 40\,MHz bandwidth for 802.11n WiFi, the time domain resolution is merely 25\,ns. This means that if the path length difference of two signals is within 7.5\,m (25\,ns $\cdot$ 3 $\cdot$ $10^8$\,m/s), the two signals cannot be separated in time domain. Fortunately, larger bandwidth is expected in future WiFi standards which can be utilized to further increase the sensing resolution. As for multi-person respiration sensing, the BBS (Blind Source Separation) or FFT (Fast Fourier Transformation) techniques could also be utilized to separate a set of source signals with different frequencies, as demonstrated in \cite{wang2016human, wang2017phasebeat, yue2018extracting}. 
	
The other possible solution is that we can employ antenna array (multiple antennas) to separate the signals coming from different directions in space domain. Actually, the proposed CSI-ratio model can be combined with the multiple-antenna techniques to increase both the sensing range and resolution. On top of the multiple-antenna setting, advanced signal processing techniques such as MUSIC (MUltiple SIgnal Classification) and multi-dimensional information fusion \cite{xie2018md} can be utilized to further increase the signal separation resolution.

\subsection{Commodity WiFi NICs with Multiple Antennas}
\label{sec:discussions:3}
In this work, we propose to utilize the ratio of CSI readings from two antennas at the WiFi receiver for respiration sensing.
We implement our FarSense system with commodity Intel 5300 NICs equipped with one antenna at the transmitter and two antennas at the receiver.
Actually, FarSense can work with other multi-antenna commodity WiFi NICs as long as the NICs can provide us CSI information.

\begin{itemize}
	\item FarSense does not depend on any specific MIMO mechanism (precoding, beamforming, multiplexing, etc) to work.
	The MIMO mechanisms do affect the data transmission but have very little effect on human sensing with CSI. Thus, no matter what MIMO mechanisms are adopted at the WiFi transceivers, FarSense works as long as the CSI streams from two receiving antennas share the same clock, which is true for current commodity MIMO WiFi NICs.
	To obtain the CSI data, commodity MIMO WiFi NICs such as Intel 5300 \cite{halperin2011tool} and Atheros AR9580 \cite{xie2018swan} can be  used (with three antennas supported).
	With more and more wireless sensing applications proposed, we believe more commodity MIMO WiFi NICs may support exportation of the CSI information for sensing in the future.
	\item Assuming that we have $M$ transmitting antennas and $N$ receiving antennas, we are able to obtain $MN$ CSI readings at each timestamp for every sub-carrier. By selecting one CSI reading as the numerator of CSI ratio and another one as the denominator, we will get $A_{MN}^2=MN(MN-1)$ CSI ratios for human sensing. Since different antenna pairs experience different multipath and environment noise, we can then employ different pairs of antennas to obtain the CSI ratio and exploit the complementarity of those pairs for better sensing performance. We plan to explore the MIMO sensing capability using the CSI-ratio model in our future work.
\end{itemize}
\section{Conclusion} \label{sec:conclusion} 
Contactless respiration sensing with pervasive WiFi signals is promising for real-life deployment. However, the small sensing range of existing systems severely limits their application in reality.  
In this paper, we propose to employ the ratio of CSI readings from two adjacent antennas to significantly push the sensing range limit from room level~(2-4m) to house level (8-9m).
We also propose novel methods to elaborately combine the complementary amplitude and phase for sensing in a much finer-grained way, further improving the sensing accuracy and range. With the proposed methods, for the first time, we are able to enable through-wall respiration sensing with commodity WiFi hardware. We believe the proposed system  greatly reduces the gap between lab prototype and real-life deployment. 
In addition, the proposed ratio of CSI readings from two adjacent antennas could be used as a new base signal in all MIMO devices, which could help to achieve high SNR and benefit a large group of other sensing applications.   

\begin{acks}

This research is supported by National Key Research and Development Plan under Grant No.2016YFB1001200, Peking University Information Technology Institute (Tianjin Binhai).

\end{acks}

\bibliographystyle{ACM-Reference-Format}
\bibliography{ref}

\end{document}